\documentclass[a4paper,11pt]{article}
\pdfoutput=1 
\usepackage{jheppub} 
\usepackage{neuralnetwork}
\usepackage{caption}
\usepackage{subcaption}
\usepackage{multicol}
\usepackage{graphicx} 
\usepackage{cancel}
\usepackage{grffile}
\usepackage{braket}
\usepackage{booktabs}
\usepackage{rotating}
\usepackage{multirow}
\usepackage{xspace}

\usepackage[normalem]{ulem}

\usepackage[T1]{fontenc}
\preprint{ADP-20-29/T1139}

\author[a]{Melissa van Beekveld,}
\author[b]{Philip Grace,}
\author[c]{Anders Kvellestad,}

\author[b]{Adam Leinweber,}
\author[b]{and Martin White}

\affiliation[a]{Rudolf Peierls Centre for Theoretical Physics, Clarendon Laboratory, 20 Parks Road, Oxford OX1
3PU, UK}
\affiliation[b]{ARC Centre of Excellence for Dark Matter Particle Physics, University of Adelaide, North Terrace, SA 5005}
\affiliation[c]{Department of Physics, University of Oslo, N-0316 Oslo}

\emailAdd{melissa.vanbeekveld@physics.ox.ac.uk}
\emailAdd{adam.leinweber@adelaide.edu.au}
\emailAdd{martin.white@adelaide.edu.au}
\title{\boldmath Simple, but not simplified: A new approach for optimising beyond-Standard Model physics searches at the Large Hadron Collider}

\abstract{
Searches for beyond-Standard Model physics scenarios, such as supersymmetry (SUSY), at the Large Hadron Collider (LHC) are frequently optimised on simplified models. After assuming particular particle production and decay processes, analyses are optimised by tuning event selections on benchmark models generated in 2D planes of parameters, with all other parameters held fixed. Motivated by recent evidence that this removes sensitivity to a large volume of viable SUSY models, we propose an alternative approach based on dimensional reduction of global fit results. Starting from the results of a global fit of the 4D electroweak minimal supersymmetric standard model performed by the GAMBIT collaboration, we show how to define a 2D plane by using a variational autoencoder to map points in the original 4D parameter space to a 2D latent space. This allows for easy visualisation of the 4D global fit results, which generates insights into what we may be missing at the LHC. Furthermore, the invertible nature of the map to the 2D plane allows experimentalists to choose and simulate benchmark models in a 2D plane (as they could for a simplified model) whilst still accessing the full range of phenomenological scenarios identified by the global fit in the 4D model. We provide a demonstration of how our visualisation and benchmark model simulation process works, and develop an analysis that is able to exclude four benchmark models not excluded by the ATLAS and CMS collaborations in the set of results used for the global fit.}

\begin{document} 
\maketitle
\flushbottom

\section{Introduction}
\label{sec:intro}
Despite a broad program of precision measurements and direct searches for new particles at the Large Hadron Collider (LHC), no evidence for beyond-Standard Model (BSM) physics has yet been obtained. A complicating factor in particle searches is the fact that there is no unambiguous prediction of the phenomenology that we can expect to observe at the LHC. The plethora of well-motivated extensions of the Standard Model (SM) come associated with a number of free parameters, many of which are relatively unconstrained \emph{a priori}, even after considering the impact of Run 2 LHC data. This means that any choice of a specific model or framework -- for example, supersymmetry (SUSY) -- really represents a huge space of possible options. Although we can usually simulate the expected LHC behaviour for any specific choice of model parameters, there is not always a clear motivation for which sets of parameters to explore in any particular analysis.

To make things worse, the signals expected from BSM models are by their very nature significantly rarer than the SM background processes that produce similar final states. This means that searches for new particles must be very heavily optimised using variables that exploit kinematic differences between the signal and background processes, leading to a strong dependence on the signal models used for that optimisation.  
Typically, such optimisations are performed using a small number (i.e.~2--3) of fundamental parameters of a given theory. This shields the rich structure a BSM model usually has, and allows many spectra to escape detection at the LHC, simply because the analysis was not optimised on them. To give a concrete example, the dominant paradigm for optimising supersymmetry searches at the LHC has been to define a series of \emph{simplified models}~\cite{Alves:2011sq,LHCNewPhysicsWorkingGroup:2011mji}, in which a particular pair of sparticles is assumed to be produced, with particular assumed decay chains. Analyses are optimised by simulating benchmark signal models chosen from parameter planes within those models, e.g.~a pair of sparticle masses assuming fixed branching ratios, and fixed values for other masses that enter the simplified model. A variety of planes are explored for fixed values of the other parameters, often with the assumption that exploring enough planes will cover most of the viable options for LHC phenomenology. Results can be used to constrain more complex models of supersymmetry by resolving the more complex model onto the basis of simplified models~\cite{Gutschow:2012pw,Kraml:2013mwa,Alguero:2021dig}, although these results are conservative compared to those obtained using the more complex model itself.

Simplified models in SUSY arose as a pragmatic compromise solution, based on the fact that available computing power severely limits the number of benchmark models that can be simulated for optimisation studies. Choosing hyperplanes in the space of sparticle masses and branching ratios allowed for a more general exploration of possible phenomenology than earlier attempts to optimise searches within the framework of minimal supergravity~\cite{PhysRevLett.49.970,BARBIERI1982343,IBANEZ198273,PhysRevD.27.2359,10.1143/PTP.70.542}. However, each simplified model plane still only represents a vanishingly thin slice of a large dimensional space of possible masses and branching ratios, and aggressive optimisation on these planes might leave little sensitivity to the models encountered once one wanders off them. Indeed, results from global fits (see e.g.~\cite{Aad:2015baa,Khachatryan:2016nvf,CMSSM,MSSM,GAMBIT:2018gjo,GAMBIT:2023yih}) indicate that large volumes of the BSM parameter space are completely unconstrained, even when the production cross sections would in principle allow for detection. 

One possible way forward is to use model-independent searches to circumvent the need for optimisation in the first place. Such approaches have existed for a long time, for example the D0 collaboration at the Tevatron have developed an unsupervised, multivariate signal detection algorithm named SLEUTH~\cite{Abbott:2000fb,Abbott:2000gx,Abbott:2001ke, Abazov:2011ma}, the H1 Collaboration~\cite{Aktas:2004pz,Aaron:2008aa} at HERA used a 1-dimensional  signal detection algorithm, 
and the CDF Collaboration~\cite{Aaltonen:2007dg, Aaltonen:2008vt} at the Tevatron also developed a 1-dimensional signal-detection algorithm. The BUMPHUNTER algorithm has been operated in a similar vein at the LHC and the Tevatron~\cite{Choudalakis:2011qn}. More recent model-independent LHC searches have been performed by the ATLAS and CMS collaborations~\cite{Aaboud:2018ufy,CMS:2008gya,ATLAS:2017irs,Asadi:2017qon,CMS:2011fra}. Very recently, there has been a vast increase of interest in the use of machine-learning techniques to perform model-independent searches. Examples include the use of 1.~neural networks to compare observations with a set of reference events~\cite{DAgnolo:2018cun,DAgnolo:2019vbw}; 2.~(variational) autoencoders in jet substructure applications and general LHC searches~\cite{Farina:2018fyg,Heimel:2018mkt,hajer2018novelty, cerri2019variational}; and 3.~unsupervised/weakly-supervised anomaly detection techniques~\cite{Romao:2020ocr,kim2015deep,Andreassen:2020nkr,Nachman:2020lpy,Collins:2019jip,Dery:2018dqr,Collins:2018epr,Benkendorfer:2020gek}. In Ref.~\cite{Aarrestad:2021oeb} many unsupervised anomaly detection techniques were compared using a handful of BSM models. In more recent studies, various applications of (variational) autoencoders for model-agnostic searches have been explored in Refs.~\cite{Buss:2022lxw,Ngairangbam:2021yma,Karagiorgi:2021ngt,Canelli:2021aps,Chekanov:2021pus,Mikuni:2021nwn,Jawahar:2021vyu,Fraser:2021lxm,Hallin:2021wme,Govorkova:2021utb,Barron:2021btf,vanBeekveld:2020txa}. Non-neural network algorithms have also recently been explored in Refs.~\cite{Finke:2022lsu,Krzyzanska:2022mto,Alvarez:2021zje,Yang:2021kyy,Caron:2021wmq,Mullin:2019mmh}. However, the issue with fully-unsupervised searches is that they struggle to have sensitivity for signal models on the threshold of discovery, which are known to be easy to discover if one knew what they were. This is only a specific case of the general statement that a supervised approach will always outperform an unsupervised approach where applicable.

In this paper, we revisit the problem of optimising LHC searches for SUSY, in an attempt to find something more general than the simplified model approach. The starting point for our method is that it has long been possible to perform rigorous statistical global fits of relatively general supersymmetric scenarios \cite{Baltz04,Allanach06,SFitter,Ruiz06,Strege15,Fittinocoverage,Catalan:2015cna,MasterCodeMSSM10,2007NewAR..51..316T,2007JHEP...07..075R,Roszkowski09a,Martinez09,Roszkowski09b,Roszkowski10,Scott09c,BertoneLHCDD,SBCoverage,Nightmare,BertoneLHCID,IC22Methods,SuperbayesXENON100,SuperBayesGC,Buchmueller08,Buchmueller09,MasterCodemSUGRA,MasterCode11,MastercodeXENON100,MastercodeHiggs,Buchmueller:2014yva,Bagnaschi:2016afc,Bagnaschi:2016xfg,Allanach:2007qk,Abdussalam09a,Abdussalam09b,Allanach11b,Allanach11a,Farmer13,arXiv:1212.4821,Fowlie13,Kim:2013uxa,arXiv:1503.08219,arXiv:1604.02102,Han:2016gvr,Bechtle:2014yna,arXiv:1405.4289,arXiv:1402.5419,MastercodeCMSSM,arXiv:1312.5233,arXiv:1310.3045, arXiv:1309.6958,arXiv:1307.3383,arXiv:1304.5526,arXiv:1212.2886,Strege13,Gladyshev:2012xq,Kowalska:2012gs,Mastercode12b,arXiv:1207.1839,arXiv:1207.4846,Roszkowski12,SuperbayesHiggs,Fittino12,Mastercode12,arXiv:1111.6098,Fittino,Trotta08,Fittino06,arXiv:1608.02489,Mastercode15,arXiv:1506.02499,Mastercode17}. Since 2017, the open-source Global and Modular Beyond-Standard Model Inference Tool (\texttt{GAMBIT}) \cite{gambit, GUM} has made global fits of generic BSM models easier, allowing SUSY models to be explored using a wide range of collider and astrophysical data~\cite{grev,CMSSM,MSSM,GAMBIT:2018gjo,GAMBIT:2023yih}. The results of such studies provide a set of supersymmetric parameter points that remain viable given past null observations of SUSY in the wide range of covered experiments, albeit in a high-dimensional space that is hard to visualise. The problem of optimising LHC searches then boils down to how best to use this information to inform future sparticle searches. Our proposed solution is to use a dimensional reduction technique to create an invertible map from allowed points in the original SUSY parameter space to a 2D plane, in which one can generate benchmark points for search optimisation. This 2D plane of parameters encapsulates the often rich set of still-viable phenomenological scenarios identified by the global fit, rather than being a slice through the original space that ignores a bulk of interesting models.\footnote{We note that the simplified model approach is itself an example of dimensional reduction, except that the SUSY parameters are first mapped to a very large space of possible masses and branching ratios, and then most of the variables are thrown away arbitrarily to generate the 2D plane in which analyses will be optimised.} Dimensional reduction is closely related to data visualisation, and we also describe ways in which our mapping can be used to explore global fit results in an intuitive way to refine future particle searches. 

To demonstrate our approach, we consider the electroweak minimal supersymmetric SM (EWMSSM), a variant of the MSSM where all sparticles are decoupled except for the four neutralinos and two charginos. At tree level the masses, couplings, and production cross sections for these six electroweakinos are given by only four parameters: the bino and wino masses $M_1$ and $M_2$, the higgsino mass $\mu$ and the ratio between the vacuum expectation values of the Higgs doublets $\tan \beta$. It is our aim to reduce the dimensionality of the optimisation problem to two, which would allow for a simple visualisation and scanning of the parameter space. For this dimensional reduction we use a variational autoencoder (VAE), trained on the results obtained from a GAMBIT global fit of the EWMSSM that investigated the impact of ATLAS, CMS and LEP sparticle searches~\cite{GAMBIT:2018gjo}. This global fit contained up-to-date results for 2018, which limits it to results that used 36 fb$^{-1}$ of integrated luminosity at a centre of mass energy of 13 TeV. We therefore do not include searches for SUSY post-2018,
and our results serve to illustrate the method rather than advocate a new search region. We present an approach that could have extracted more from the specific dataset that informed the global fit in Ref.~\cite{GAMBIT:2018gjo}, and the future use of this approach should similarly extract more from the current data than a purely simplified model-based approach.  

We aim to achieve a useful middle ground that retains the benefit of a supervised search approach, which yields a better rejection of SM backgrounds based on a detailed knowledge of kinematic differences with the signal than unsupervised methods, but without the limitations of a narrow range of simplified models. Holes in the simplified model approach arise because viewing the model in terms of its masses and branching ratios generates a high dimensional space of possible options even though the original phenomenology might be described by only a few numbers. Although particular GUT-scale theories might be poorly motivated, weak-scale scenarios such as the EWMSSM are useful examples of relatively low-dimensional scenarios that make sense at this point in history. A Higgs mass of 125 GeV requires heavy stop quarks in the MSSM, which in turn implies a heavy coloured sector. Then one can postulate that the electroweakinos are the only superpartners accessible at the LHC in the near future. This SUSY scenario (the EWMSSM) can be reduced to a 2D \emph{simple} model for optimisation, as compared to choosing a~\emph{simplified model} which throws away a considerable volume of options. Our technique can easily be generalised to higher dimensional parent models, although it remains to be seen how successful the dimensional reduction would be in those cases.

This paper is structured as follows: in Section~\ref{sec:VAE} we provide a brief introduction to variational autoencoders for dimensional reduction. In Section~\ref{sec:gambit} we describe the process of applying a VAE to GAMBIT global fit results to define a 2D plane in which viable SUSY models can be visualised and used for benchmark studies. In Section~\ref{sec:latent-space} we demonstrate how visualisation works by examining interesting quantities in our defined 2D plane to gain insight into the kind of models that evaded the searches included in the original global fit. In Section~\ref{sec:optimisation} we select four unexcluded points from the latent space defined in the previous section, and construct an analysis that excludes each of them at the 95\% confidence level. We summarise our conclusions in Section~\ref{sec:conclusion}.

\section{Variational autoencoders for dimensional reduction}
\label{sec:VAE}
The key problem in our paper is how to take a set of points from an $n$-dimensional parameter space $\{\vec{\theta}_{\text{parent}}\}$ and map it to a set of points in a two-dimensional parameter space $\{\vec{\theta}_{\text{2D}}\}$. The $n$D parameters represent those of some parent BSM model, and we must define an invertible map that takes points from the $n$D space to their equivalent points in a 2D plane. The map must be invertible so that we can choose benchmark points in the 2D plane and perform a Monte Carlo simulation for analysis optimisation, which can only be accomplished if we know the original parent parameters for each point in the 2D plane. 

This is a canonical dimensional reduction problem, and there are a variety of techniques that can be employed to solve it. For example, an autoencoder (AE)~\cite{10.1145/1390156.1390294_autoencoder} is a special type of neural network architecture that maps a given set of input variables to the same input variables. It is designed to learn a representation of the input that typically has a lower dimensionality than that of the input. AE training is unsupervised and does not require labelling or classification. The loss function is typically chosen to be the reconstruction loss, which is
the difference between the output and the input, quantified, for example, by the mean squared
error in every dimension of the data. An AE is successful if the output and input are indistinguishable. 

The structure of an AE can be broken down into three parts: the encoder, the decoder, and the latent space that separates them. The encoder usually has multiple hidden layers with decreasing numbers of nodes, while the decoder is typically the mirror image of the encoder, having an increasing number of nodes with each layer. It is important for the number of nodes to decrease in each sequential layer of the encoder so that the network does not simply learn the identity function. At the centre of the network, between the encoder and decoder, is the latent space which is a low dimensional layer containing an abstract compressed version of the input. The dimensionality of the latent space greatly affects the performance of the network. If the dimensionality of the latent space is too low, it is possible that the network loses too much information to be able to reliably reconstruct inputs. Conversely, if the dimensionality is too high, one loses the benefits of dimensional reduction. 

\begin{figure}[t]
\centering
 \includegraphics[width=0.8\textwidth]{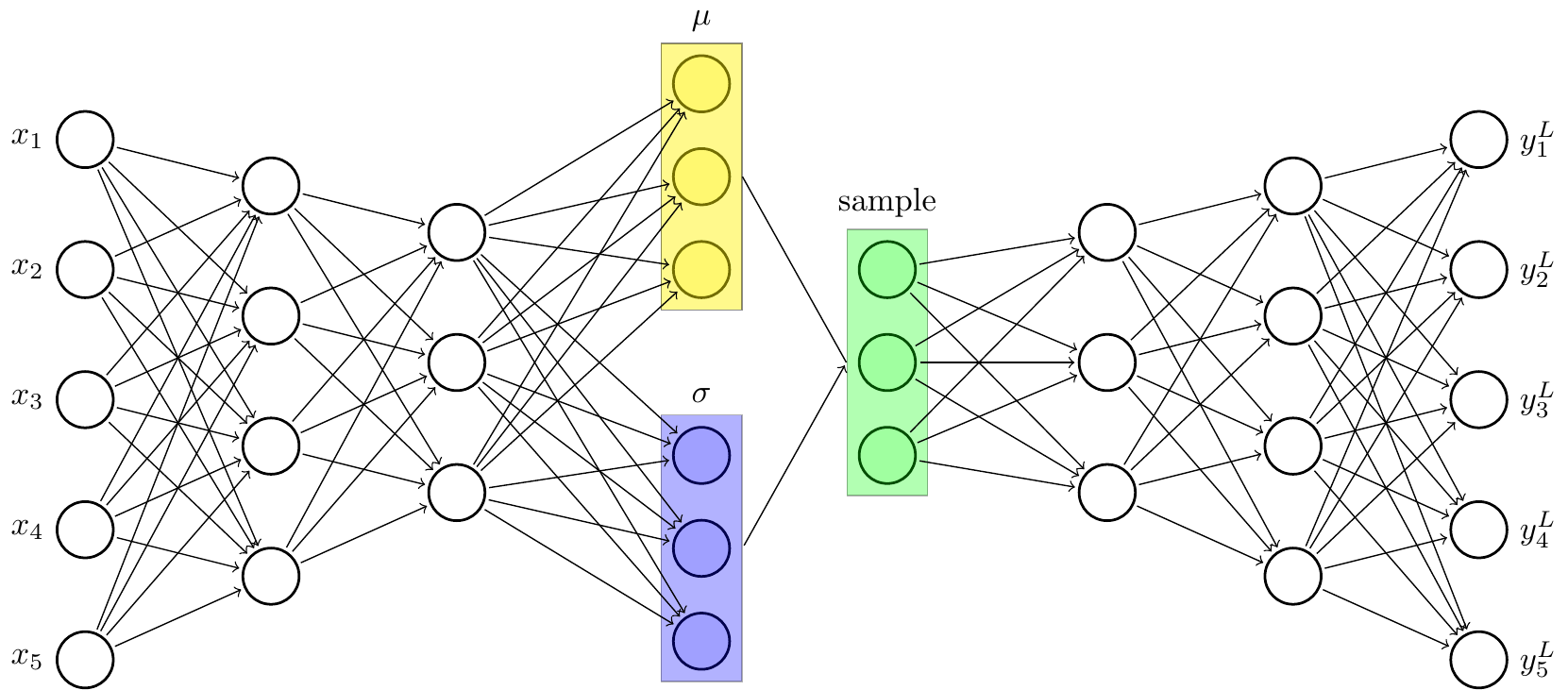}
 \caption{Schematic of a VAE where $x_i$ is the $i$th input variable and $y^L_i$ is the $i$th reconstructed value of $x_i$. In this example, the VAE has a three-dimensional latent space.}
 \label{fig:VAE}
\end{figure}

A known problem with autoencoders is that they do not exhibit ordering in the latent space. In other words, points close together in the latent space could have an arbitrary separation in the space of input variables, which is highly undesirable for our use case. The variational autoencoder \cite{kingma2013autoencoding_vae} (VAE) enforces ordering by modifying the middle part of the network. The encoder outputs two numbers per latent space dimension that represent the mean and standard deviation of a Gaussian distribution for each dimension. The decoder calculates its estimate of the original input by taking a random sample of the distributions, and decoding those. An additional term is added to the loss function so that the KL divergence~\cite{Kullback59} of these Gaussians and a standard normal distribution should be as low as possible. The mean-squared error term exists to ensure optimal reconstruction of the input variables, whilst the KL-divergence term enforces ordering in the latent space by ensuring that all inputs should be encoded as close to $\vec{0}$ in the latent space as possible. Obtaining reasonable behaviour from a given VAE architecture requires careful tuning of the relative strength of the two contributions to the loss function (also see Ref.~\cite{vanBeekveld:2020txa}). The structure of a VAE consisting of an encoder with 2 layers, the latent space, and a decoder with 2 layers is shown in Figure~\ref{fig:VAE}. 

% Describe the dataset
\section{VAE training on GAMBIT global fit results}
\label{sec:gambit}
The data used to train and test our VAE originate from a global fit of the EWMSSM performed by the GAMBIT collaboration~\cite{GAMBIT:2018gjo}. In this case, $\vec{\theta}_{\text{parent}}\equiv (M_1, M_2, \mu,\text{tan}\beta)$. After electroweak symmetry breaking, these four parameters describe the tree-level masses and couplings of four neutral and two charged mass eigenstates: the neutralinos $\widetilde{\chi}^0_i$ with $i = 1,\dots,4$, ordered in mass, and the charginos $\widetilde{\chi}^\pm_i$ with $i = 1,2$. The study assumes that the other sparticles of the MSSM are heavy and decoupled. To this end, the pseudo-scalar Higgs mass $m_A$ and the gluino mass parameter $M_3$ are set to $5$~TeV, all trilinear couplings are set to zero, and all other mass parameters except for those describing the electroweakino sector are set to $M_{\rm SUSY} = 3$~TeV. Parameter points were assigned a log-likelihood based on the searches for electroweakinos at the LEP and LHC colliders, plus constraints on the invisible widths of the $Z$ and SM-like Higgs boson. Further details on the scanning of the parameter space and the implemented searches are given in Ref.~\cite{GAMBIT:2018gjo}. 

\begin{figure}[!ht]
	\centering
	\begin{subfigure}{0.49\textwidth}
		\includegraphics[width=\textwidth]{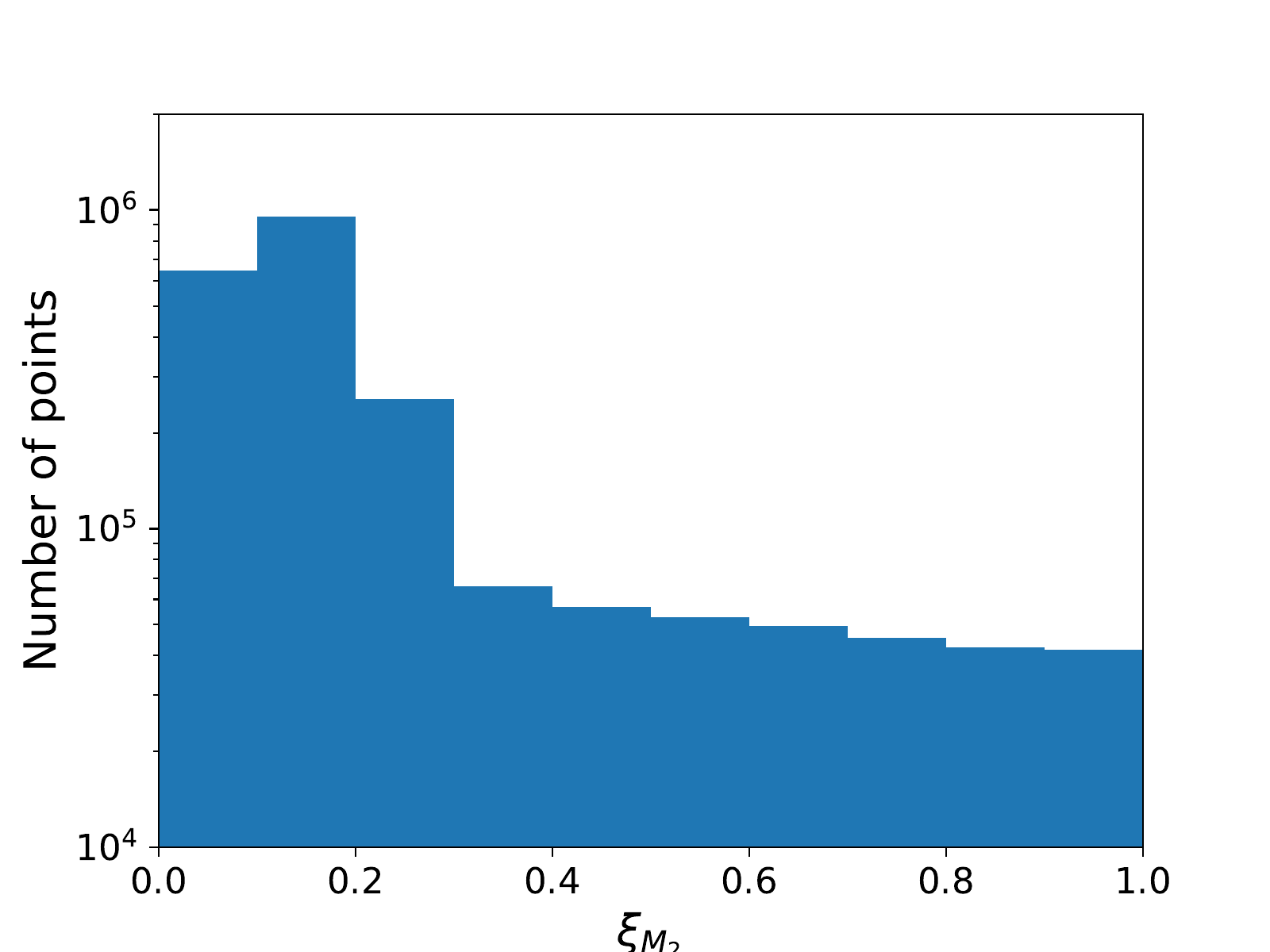}
	\end{subfigure}
	\hfill
	\begin{subfigure}{0.49\textwidth}
		\includegraphics[width=\textwidth]{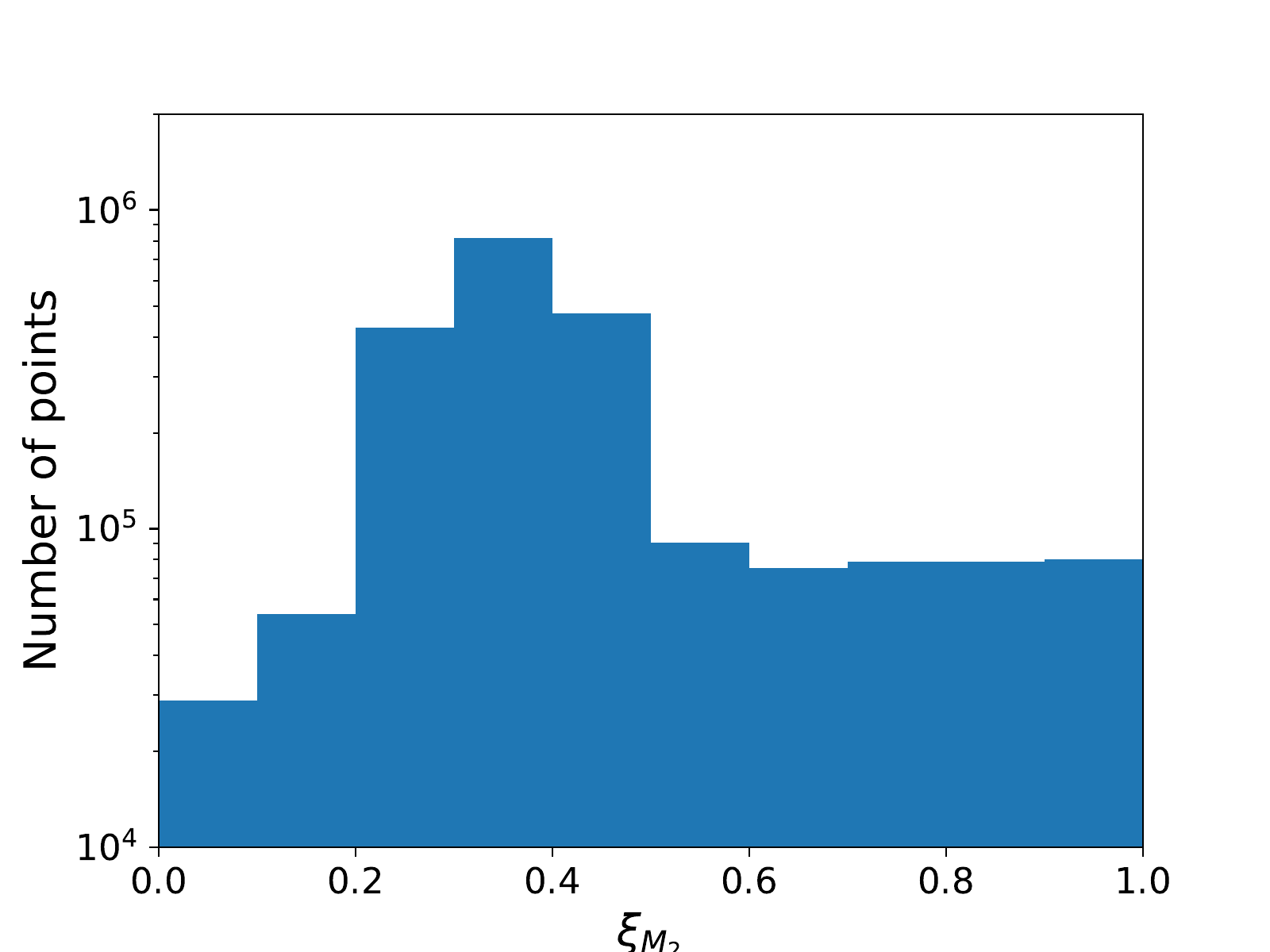}
	\end{subfigure}
	\caption{\label{fig:M2_unskew} Values of $\xi_{M_2}$ displayed as a histogram before (left) and after (right) taking the square root in order to unskew the distribution.}
\end{figure}

The full dataset we use to train and test the VAE consists of all parameter points from Ref.~\cite{GAMBIT:2018gjo} allowed at the 3$\sigma$ level when compared to the background-only expectation.\footnote{In the terminology of Ref.~\cite{GAMBIT:2018gjo} we select points based on the ``capped likelihood'' results, i.e.\ using confidence regions based on a comparison to SM expectations rather than a comparison to the best-fit EWMSSM point in the global fit.} For the visualisations and the selection of benchmark points we use the subset of points allowed at the 2$\sigma$ level. The dataset is divided into training and testing sets with a ratio of 80/20. Each of the model parameters is scaled to be between 0 and 1, i.e.~each parameter $x \in [M_1, M_2, \mu, \tan\beta]$ is transformed according to 
\begin{equation}
    \xi_{x} \equiv \frac{x-x_{\rm min}}{x_{\rm max} - x_{\rm min}}, 
\end{equation}
where $x_{\rm max}$ and $x_{\rm min}$ are the maximum and minimum values of the parameter $x$ within the dataset. These maxima and minima are detailed in Table~\ref{tab:max_min}. After scaling, the square root of $\xi_{M_2}$ is taken to reduce the negative skew of the variables' distribution. Training the VAE on $\sqrt{\xi_{M_2}}$ rather than $\xi_{M_2}$ resulted in a significant improvement in reconstruction quality. Figure~\ref{fig:M2_unskew} shows the distribution of this variable before and after unskewing.

\begin{table}[t]
	\begin{center}
		\begin{tabular}{c|c c}
			Parameter & Minimum & Maximum \\
			\hline
			$M_1$~[GeV] & -2000 & 2000\\
			$M_2$~[GeV] & 0 & 2000 \\
			$\mu$~[GeV] & -2000 & 2000 \\
			$\tan\beta$ & 1 & 70 \\  
		\end{tabular}
		\caption{\label{tab:max_min} Maximum and minimum values of each parameter before scaling.}
	\end{center}
\end{table}

To optimise the VAE, we tested the following hyperparameters:
\begin{itemize}
    \item The number of hidden layers in the encoder/decoder ranging from 3-5 with numbers of nodes between 4 and 256;
    \item Various activation functions for each of the nodes: the hyperbolic tangent ($\tanh$), the rectified linear unit (ReLU), and the sigmoid linear unit;
    \item Various loss functions:  mean-squared error, absolute error, and the $\beta$-VAE loss function with $\beta=[10^{-1},10^{-2},10^{-3},10^{-4},10^{-5},10^{-6}]$. 
\end{itemize}
 For the representation of the final results, we chose the VAE with simultaneously the lowest mean squared error (MSE) and the highest Pearson correlation coefficient between the input and the reconstructed output when run on the test set.
 This VAE is defined with a total of 11 hidden layers (5 in the encoder, 5 in the decoder, and 1 in the latent space). The number of nodes in each layer of the encoder is 100, 100, 50, 25, and 10 respectively. This form is mirrored for the decoder. The latent space is chosen to have 2 nodes to match the requirement of a 2D plane for analysis optimisation, and the latent space variables are the $\vec{\theta}_{\text{2D}} \equiv (\theta_1, \theta_2)$ that will be used for low-dimensional visualisation and optimisation. Each layer uses a $\tanh$ activation function, except for the latent space which uses the linear activation function. The loss function comparing a single point $x$ to its reconstructed counterpart $\hat{x}$ is defined as 
 \begin{equation}
     \mathcal{L} = (1-\beta)(x - \hat{x})^2 + \beta\sum_i^d\rm{KL}(\mathcal{N}(\hat{\mu}_i,\hat{\sigma}_i),\mathcal{N}(0,1))\,,
 \end{equation}
 where $\mu_i$ and $\sigma_i$ are the mean and standard deviations of the $i$th Gaussian within the $d$-dimensional latent space. The $\beta$ term is set to $10^{-6}$ and balances the relative importance of the mean squared error term and the KL divergence term. It was found that a small value of $\beta$ gave the best reconstruction quality, while still ensuring ordering within the latent space. This VAE is trained to compress and reconstruct the four transformed MSSM model parameters: $\xi_{M_1}$, $\sqrt{\xi_{M_2}}$, $\xi_{\mu}$, and $\xi_{\tan\beta}$. Training is performed over 10 iterations, with the learning rate beginning at $10^{-3}$, and halving after every iteration. The batch size is set to $1000$, and the number of epochs is $10^4$. However, if the loss on the validation set does not improve for 50 epochs, the iteration ends early. The model is only saved on epochs where the validation loss improves to avoid overfitting. 

To assess the reconstruction quality of the VAE, we calculate the Pearson correlation coefficient between each parameter and its reconstructed value. A value of $\pm1$ indicates perfect positive/negative correlation, while a value of 0 indicates no correlation. Table~\ref{tab:vae_correlation} shows these Pearson correlation coefficients. We can see that each variable has a correlation coefficient above 0.9 indicating a very strong correlation. This shows that the VAE is reconstructing the original model parameters well.

\begin{table}[t]
	\begin{center}
		\begin{tabular}{c|c}
			Parameter & Pearson Correlation Coefficient \\
			\hline
			$\xi_{M_1}$ & 0.947\\
			$\sqrt{\xi_{M_2}}$ & 0.936 \\
			$\xi_{\mu}$ & 0.927 \\
			$\xi_{\tan\beta}$ & 0.998 \\  
		\end{tabular}
		\caption{\label{tab:vae_correlation}Pearson correlation coefficients between the input parameters and their associated reconstructed output parameters. A value of $\pm1$ implies perfect positive/negative correlation, while a value of 0 implies no correlation whatsoever. A value as close as possible to $+1$ is desirable.}
	\end{center}
\end{table}

It is important to note how the distribution of models in the original space affects the training of the VAE. If a certain region of the parameter space is not well represented by the training set, it is not to be expected that the VAE will learn how to compress and reconstruct models from that region in the original space. The models used in the training of this VAE are selected by various scan settings in the original GAMBIT study, as well as the $3\sigma$ contour selection. This means that the relative density of points in the original space has no statistical meaning and is simply an artifact of the various selection processes. Since a VAE yields better performance across the original space when trained on a flat distribution, it is possible that a different set of selections would result in a training set that yields better reconstruction quality across the original space.

\section{Visualisation of the two-dimensional latent space}
\label{sec:latent-space}
After the VAE is trained, we can map any points $(M_1,M_2,\mu,\text{tan}\beta)$ to the low-dimensional parameters $(\theta_1,\theta_2)$. Starting from our GAMBIT sample of EWMSSM points that are not excluded by searches for electroweakinos using data up to 2018, we can map them to the $\vec{\theta}_{\text{2D}}$ plane and make a series of colour maps that shed light on the properties of points that evaded detection in the searches included in the original global fit. This provides a neat way to interpret multidimensional global fit results in order to determine what we are missing at the LHC. 

\begin{figure}[t]
	\centering
	\begin{subfigure}{0.49\textwidth}
		\includegraphics[width=\textwidth]{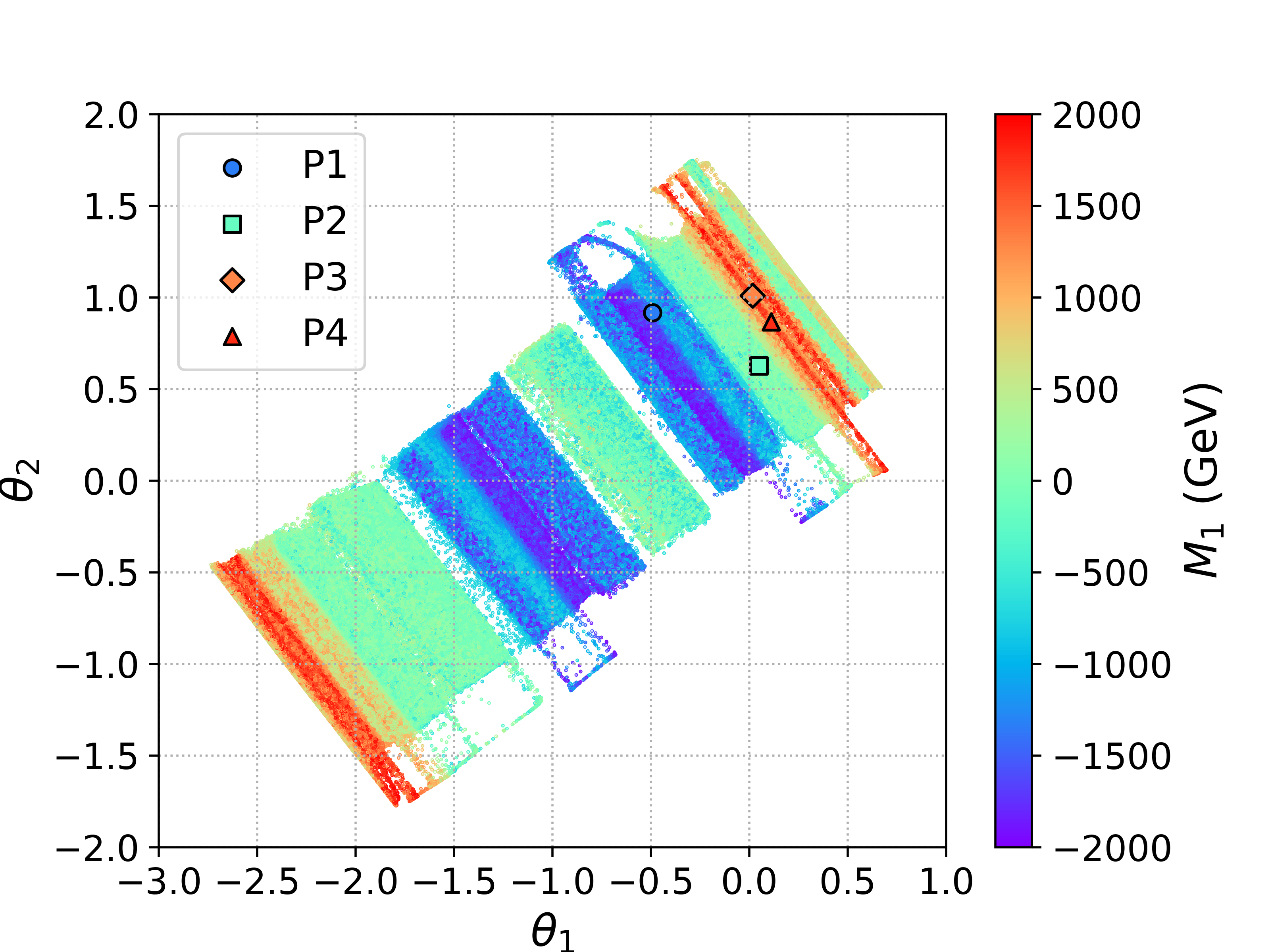}
	\end{subfigure}
	\hfill
	\begin{subfigure}{0.49\textwidth}
		\includegraphics[width=\textwidth]{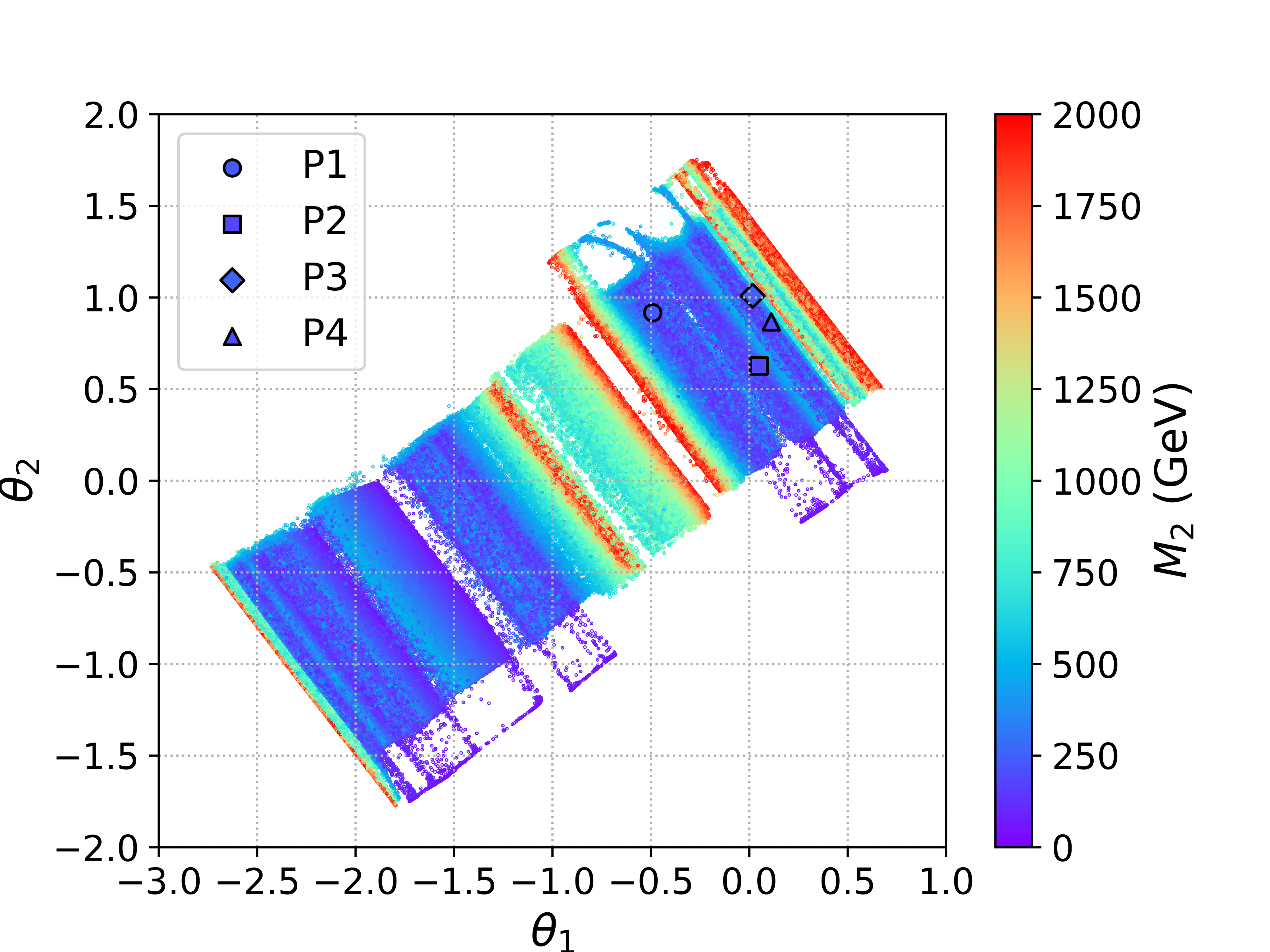}
	\end{subfigure}
	\hfill
	\begin{subfigure}{0.49\textwidth}
		\includegraphics[width=\textwidth]{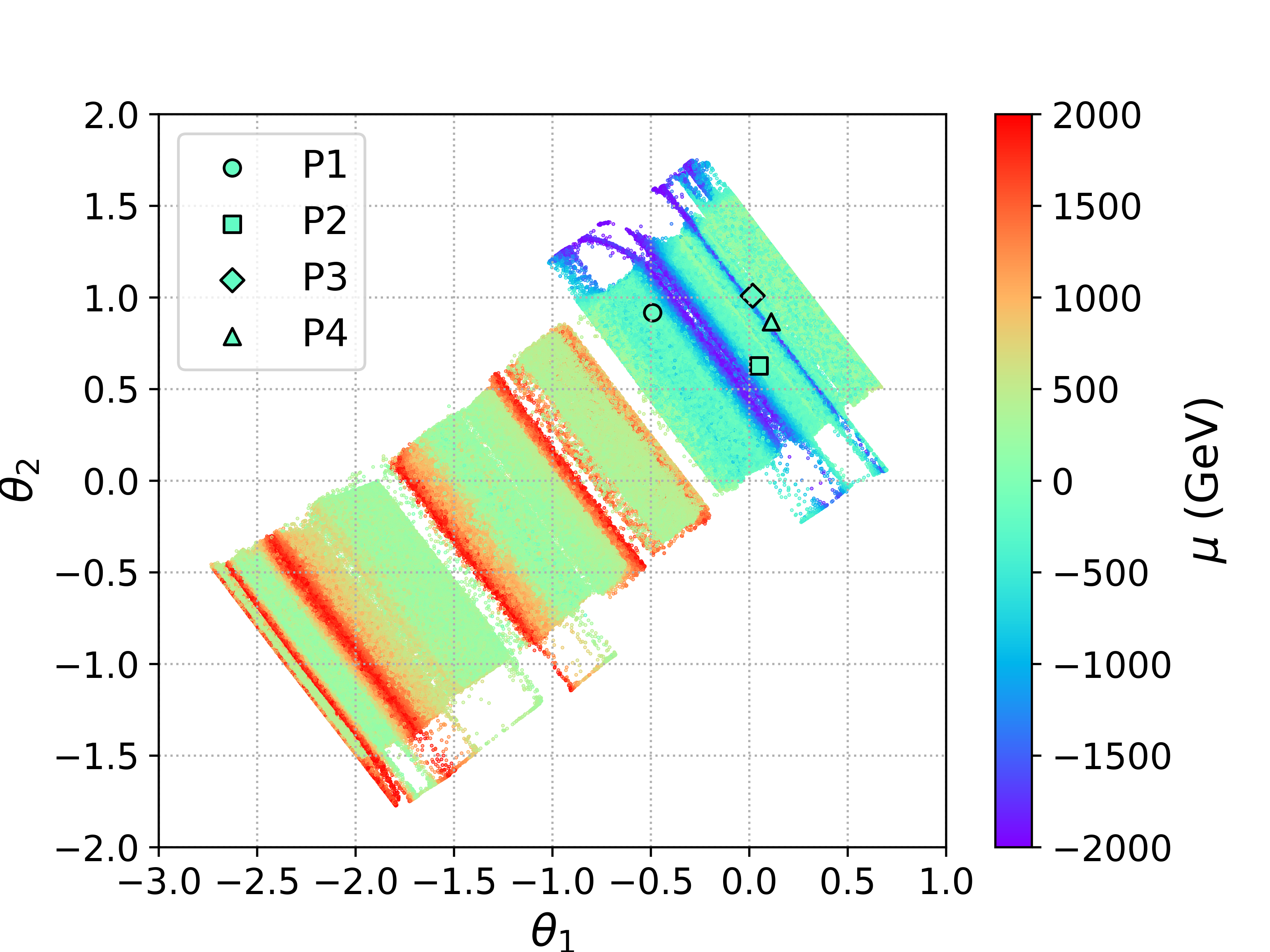}
	\end{subfigure}
	\hfill	
	\begin{subfigure}{0.49\textwidth}
		\includegraphics[width=\textwidth]{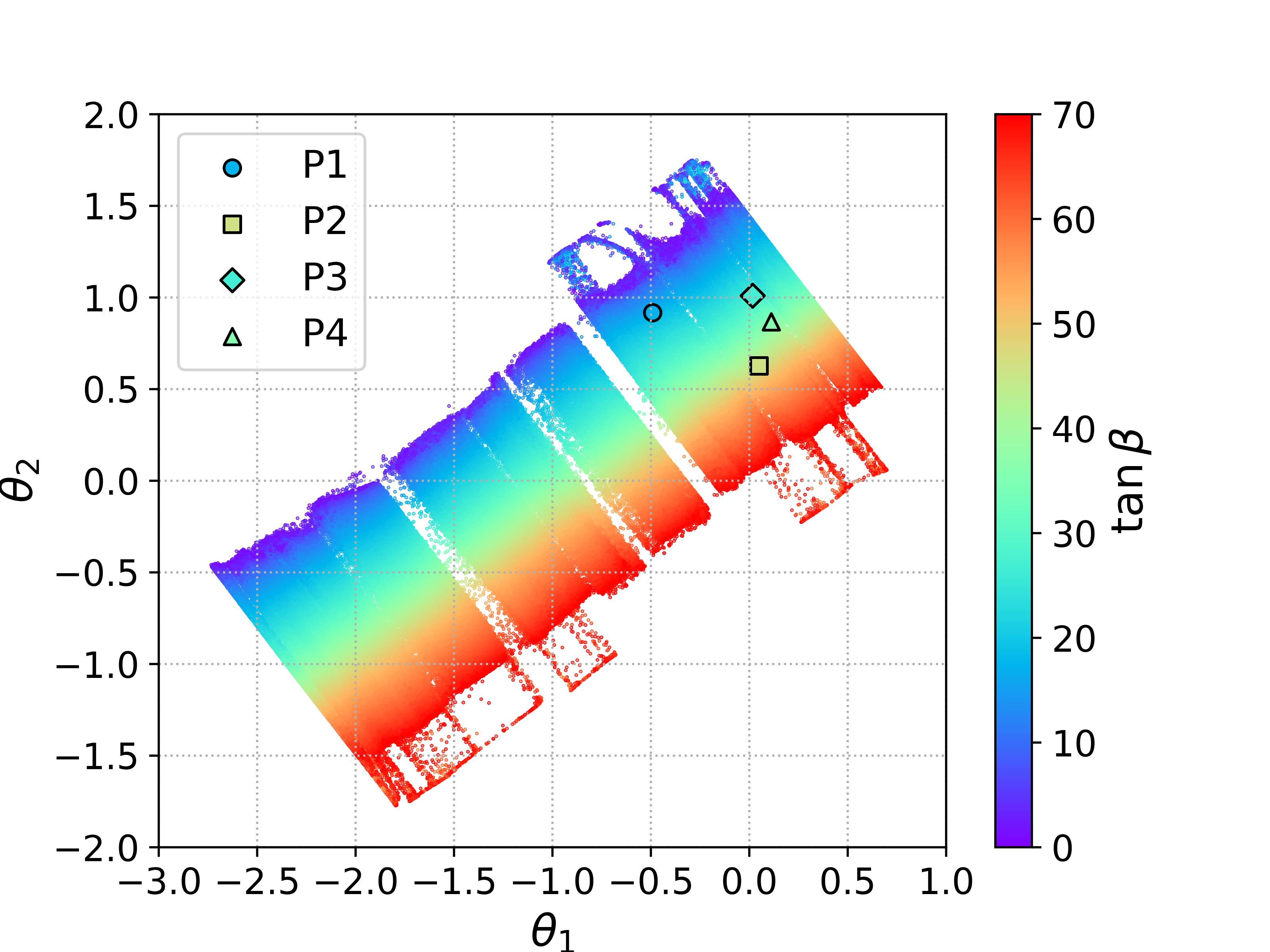}
	\end{subfigure}
	\hfill	
	\caption{\label{fig:vae_MSSM} Values of the EWMSSM input variables $M_1$ (top left), $M_2$ (top right), $\mu$ (bottom left) and $\tan\beta$ (bottom right) in the $\vec{\theta}_{\text{2D}}$ plane.}
\end{figure}

To start, we show in Figure~\ref{fig:vae_MSSM} colour maps of each of the transformed EWMSSM input variables in the $\vec{\theta}_{\text{2D}}$ plane. It is reassuring to note that each of our original input variables has continuous regions of similar colour, indicating that there is ordering in the latent space as expected for a VAE. In other words, adjacent points in the original space map well to adjacent points in the latent space, which makes optimisation and visualisation well-defined. To aid the visualisation of these variables, approximately 3\% of the most poorly reconstructed values are removed from all of the following scatter plots. The metric used to reject poorly reconstructed points is defined as 
\begin{equation}
    \alpha = \sum_i^n(x_i - \hat{x}_i)^2,
\end{equation}
where $n$ refers to each input parameter. Any point with $\alpha > 0.05$ is removed. This improves visualisation, but note that it also means that any choice of benchmark point in the latent space will not occur in a region that is poorly reconstructed by the VAE. 
We note also the presence of distinct features in each plot -- clearly the entire $\vec{\theta}_{\text{2D}}$ plane is not covered by the models in our original 4D space. The 3$\sigma$ region from the GAMBIT global fit contains a wide range of different scenarios, from scenarios where all six electroweakinos have masses below a few hundred GeV, to scenarios where the electroweakinos are largely decoupled. In terms of the EWMSSM parameters, the predicted collider phenomenology is largely determined by the magnitudes and relative ordering of the mass parameters $(M_1, M_2, \mu)$, while the value of $\tan \beta$ plays a relatively minor role. The general pattern of the VAE transformation in Fig.~\ref{fig:vae_MSSM} is that different $(M_1, M_2, \mu)$ scenarios are separated along the bottom-left to top-right direction, while different $\tan \beta$ values are distributed in the top-left to bottom-right direction.

\begin{figure}[t]
	\centering
	\begin{subfigure}{0.49\textwidth}
		\includegraphics[width=\textwidth]{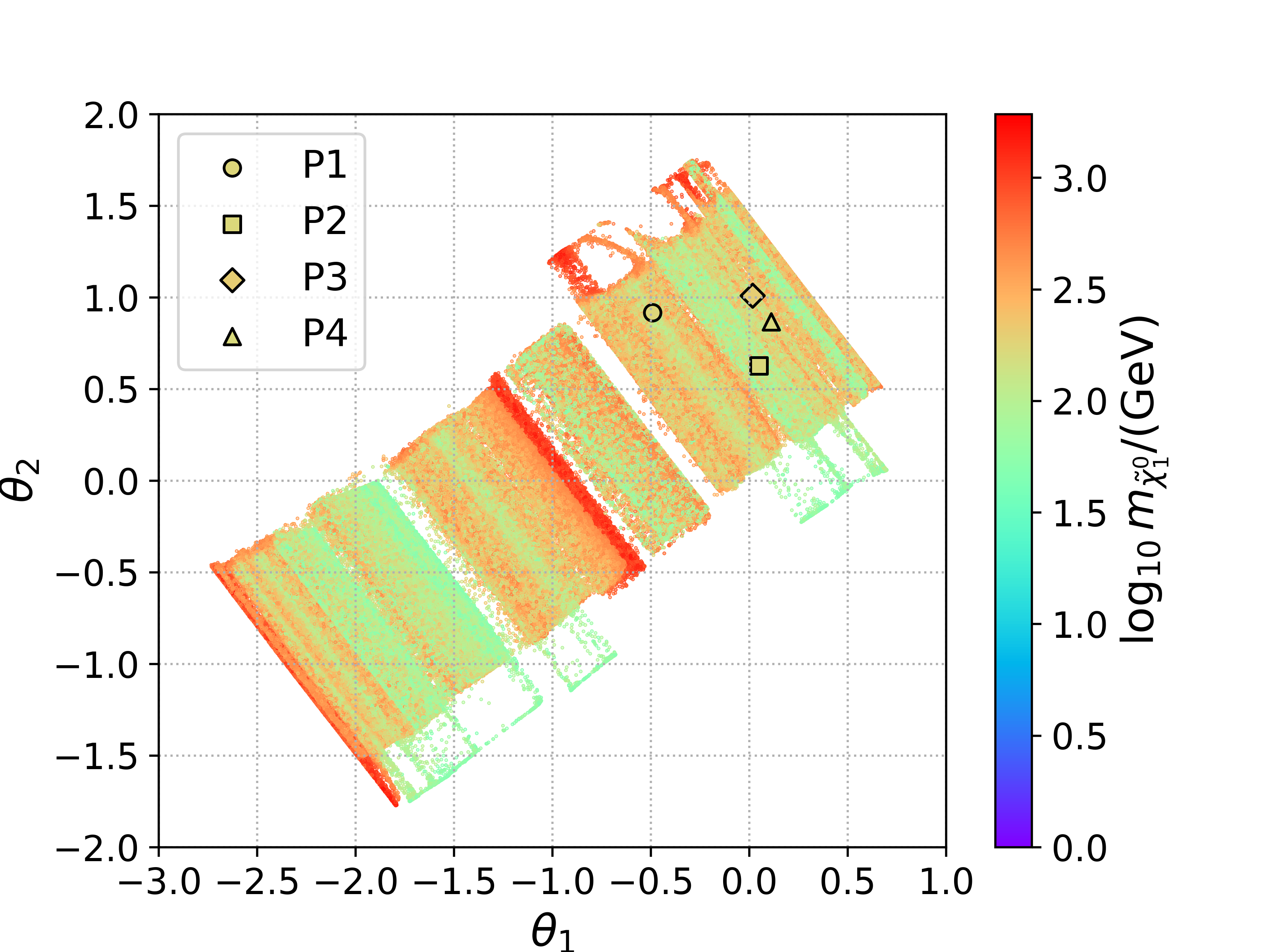}
	\end{subfigure}
	\hfill
	\begin{subfigure}{0.49\textwidth}
		\includegraphics[width=\textwidth]{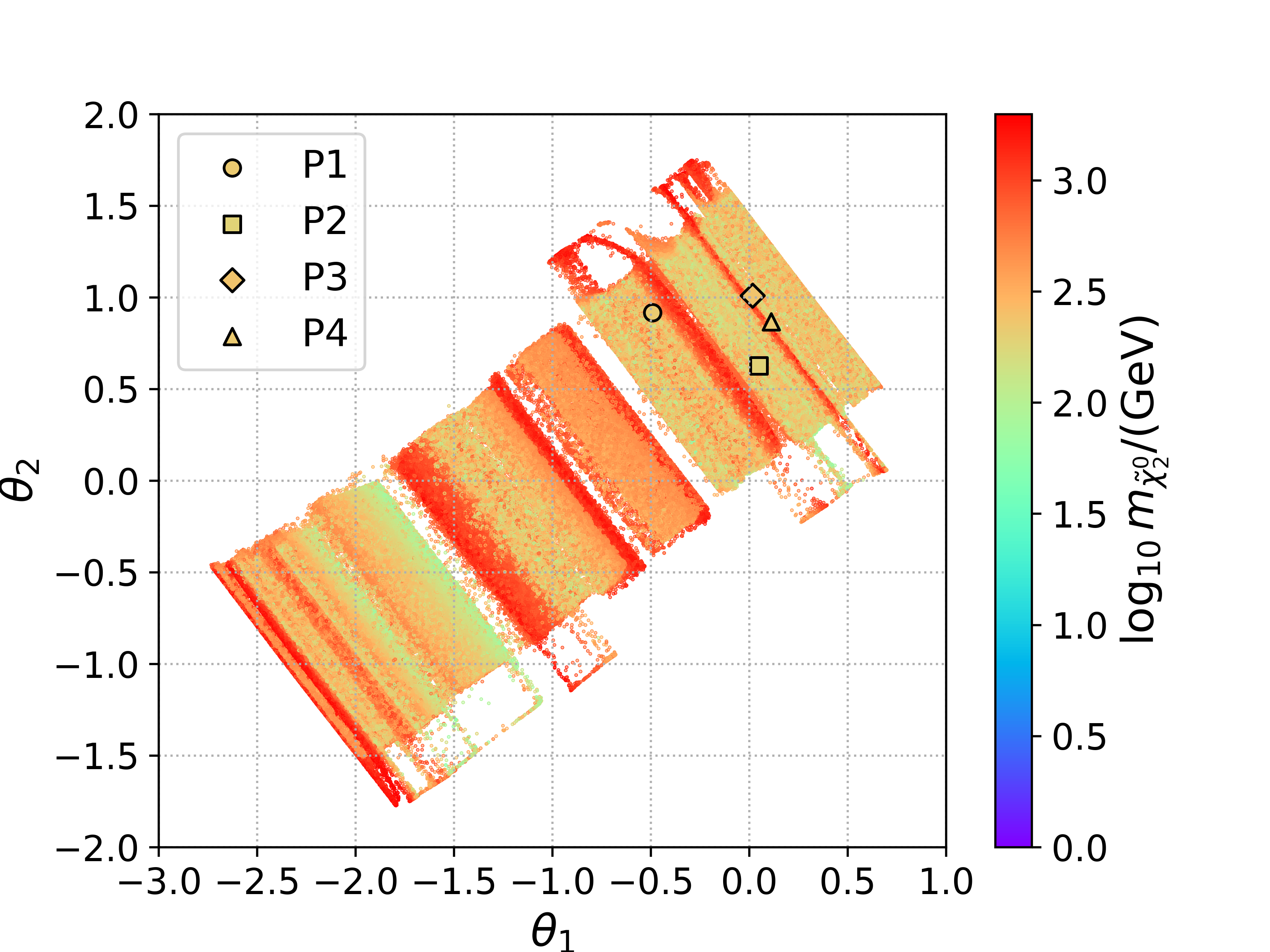}
	\end{subfigure}
	\hfill
	\begin{subfigure}{0.49\textwidth}
		\includegraphics[width=\textwidth]{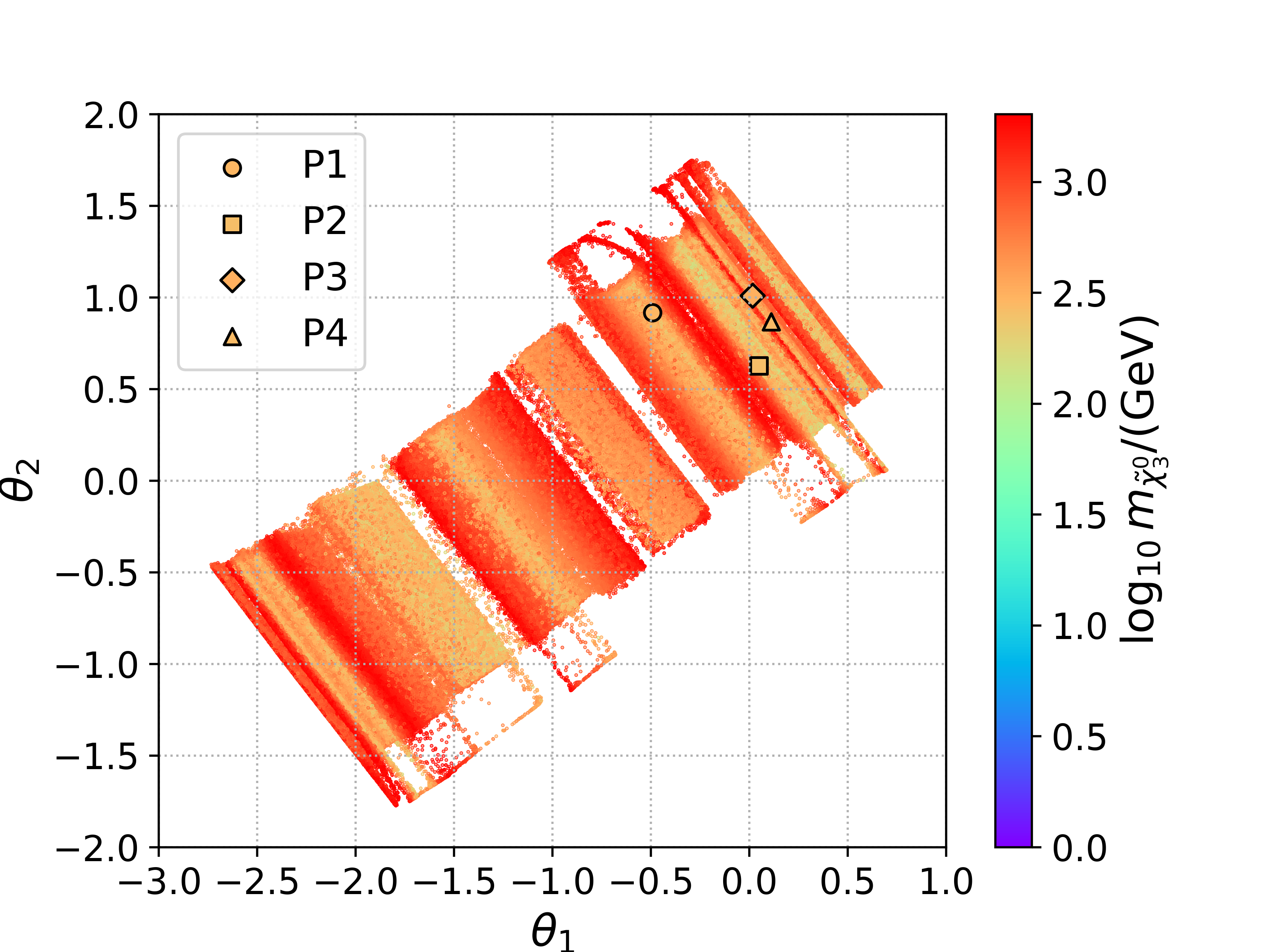}
	\end{subfigure}
	\hfill	
	\begin{subfigure}{0.49\textwidth}
		\includegraphics[width=\textwidth]{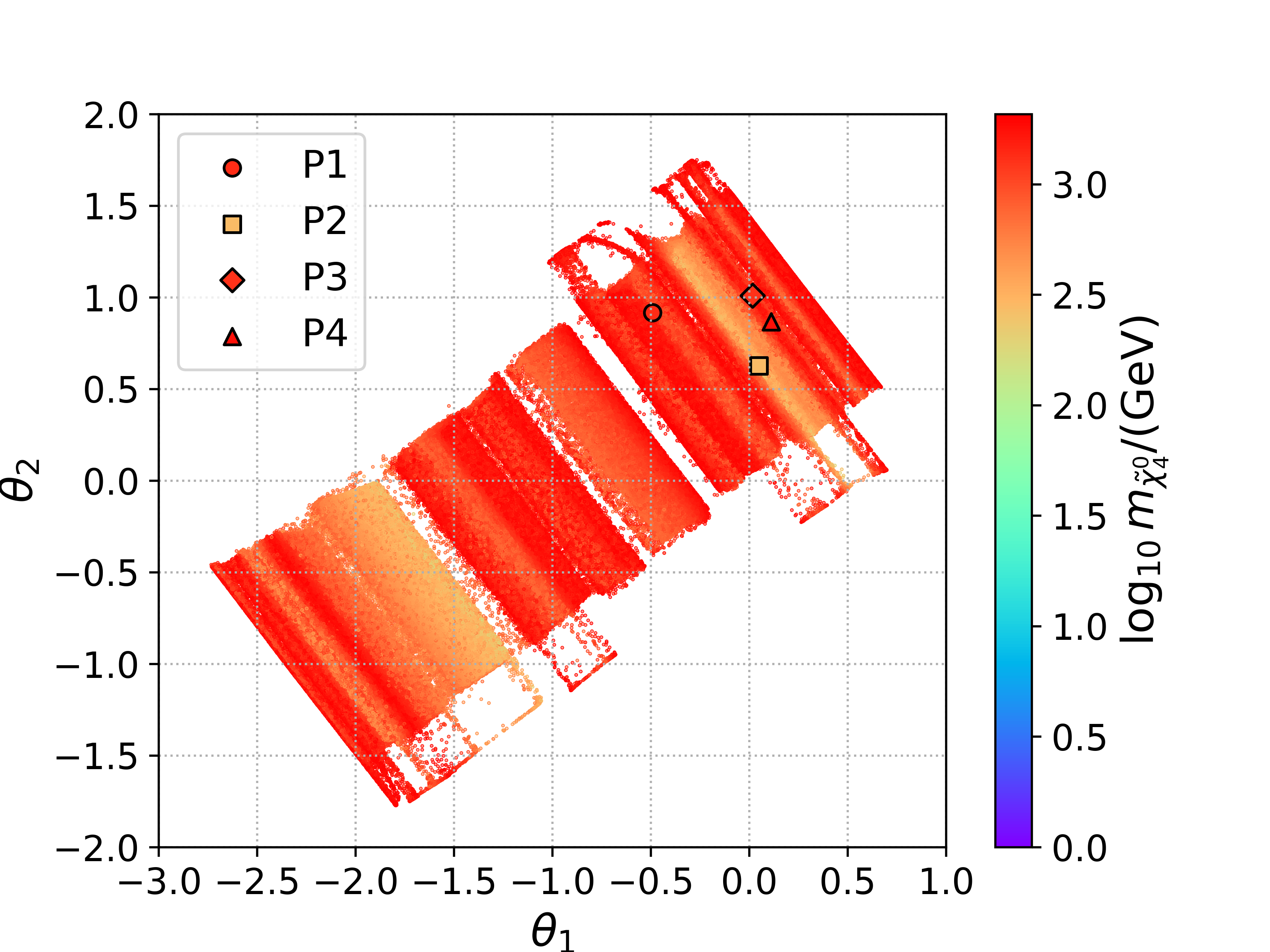}
	\end{subfigure}
	\hfill	
		\begin{subfigure}{0.49\textwidth}
		\includegraphics[width=\textwidth]{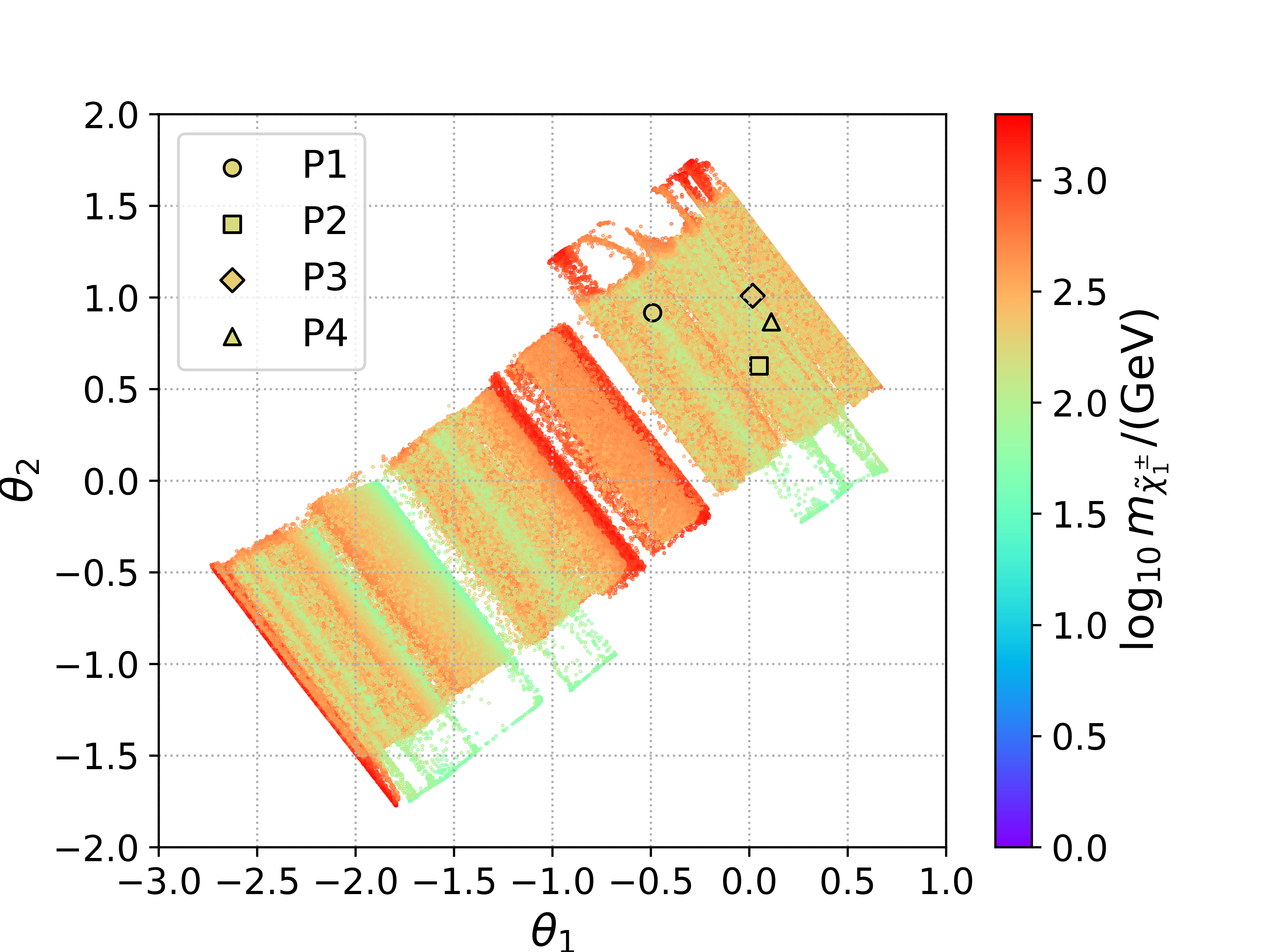}
	\end{subfigure}
	\hfill
	\begin{subfigure}{0.49\textwidth}
		\includegraphics[width=\textwidth]{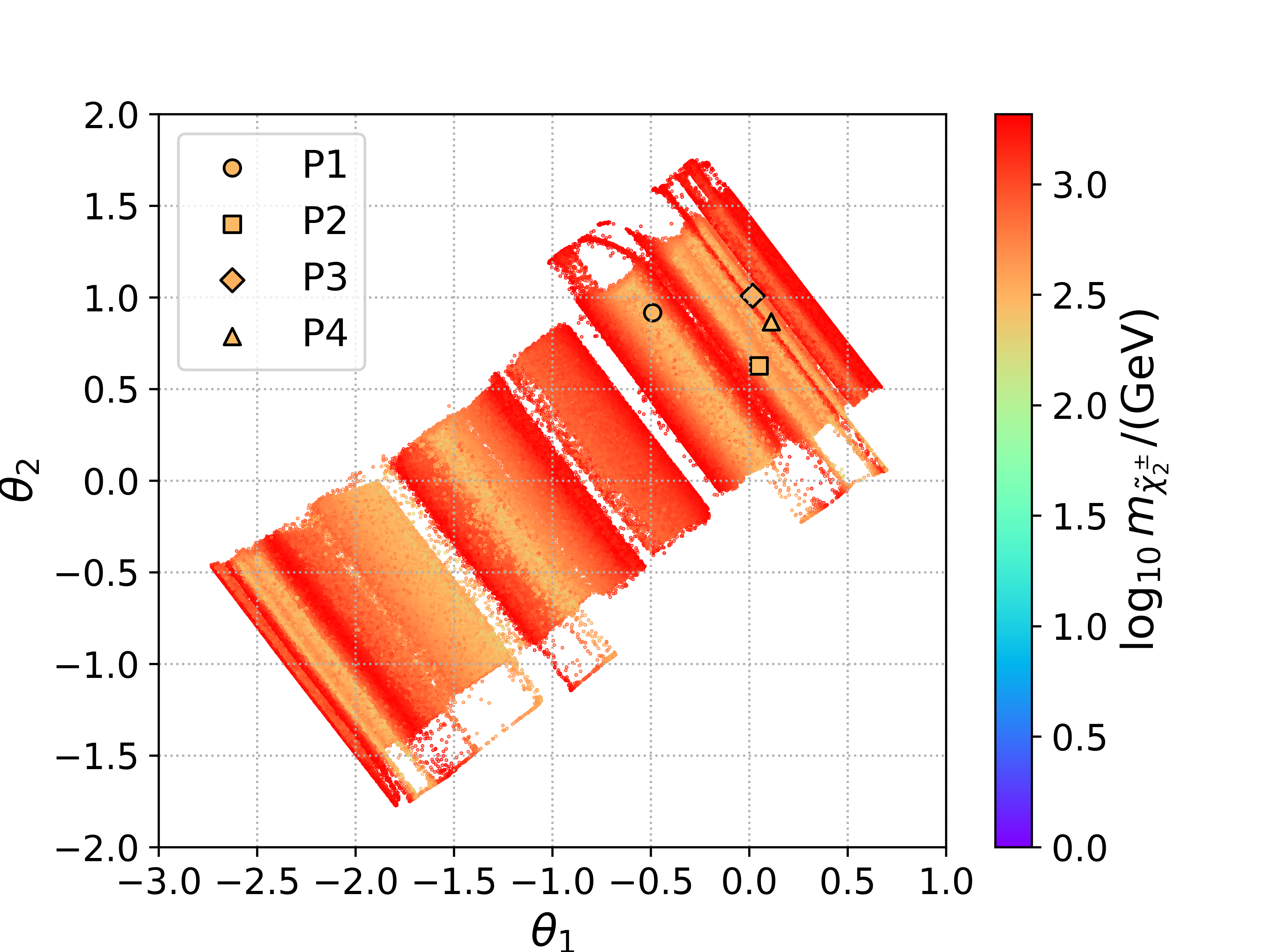}
	\end{subfigure}
	\hfill
	\caption{\label{fig:vae_MSSM_masses} The electroweakino masses in the $\vec{\theta}_{\text{2D}}$ plane.  The four benchmark points P1, P2, P3, and P4 are detailed in Section~\ref{sec:optimisation}. }
\end{figure}

We can learn more about the structures in the latent space plane by examining other quantities. In Figure~\ref{fig:vae_MSSM_masses}, we show the mass of each electroweakino in the $\vec{\theta}_{\text{2D}}$ plane. Using these figures we can identify regions that are out of reach of current experiments, but can also serve to target unprobed areas that are likely to be within reach. For example, areas with very high values for $m_{\widetilde{\chi}^0_1}$ and $m_{\widetilde{\chi}^{\pm}_1}$ are likely to be impossible to exclude currently.

\begin{figure}[t]
	\centering
	\mbox{\includegraphics[width=.5\textwidth]{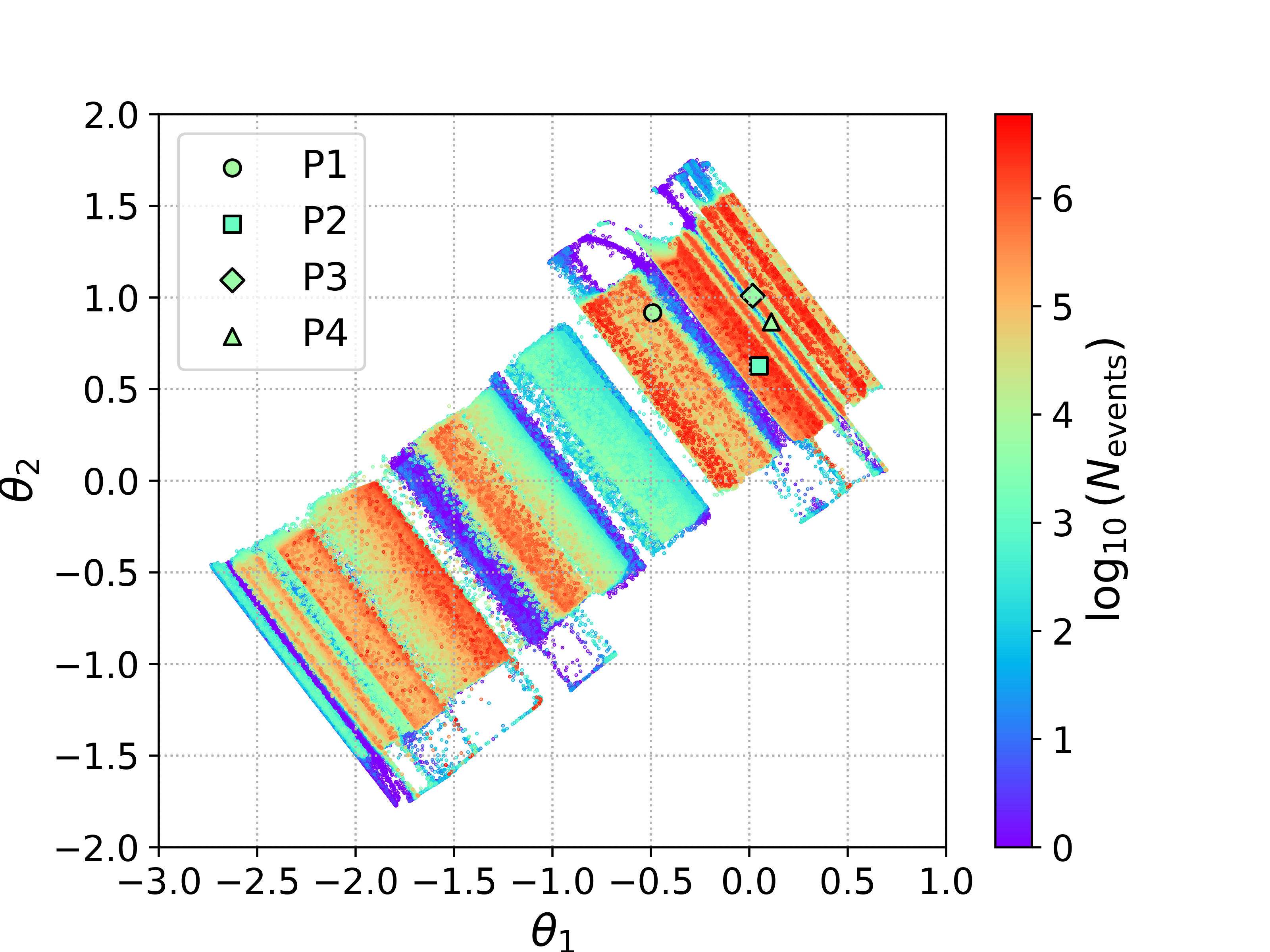} 
	\includegraphics[width=.5\textwidth]{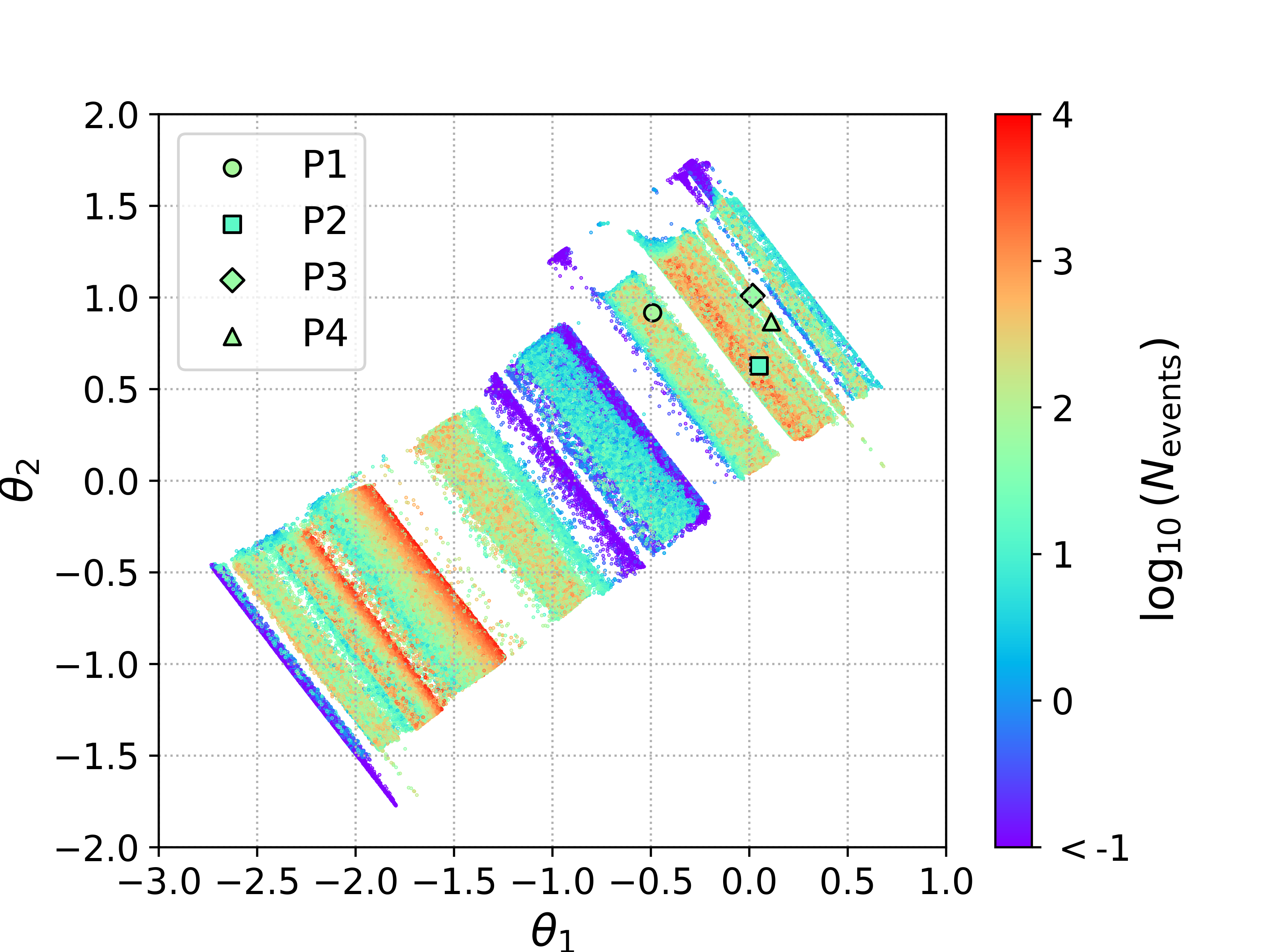}}
	\caption{\label{fig:latent_x1n2} Left: latent-space representations of points with the number of $\widetilde{\chi}_1^\pm \widetilde{\chi}_2^0$ events at 36~fb$^{-1}$ as a colour map. Right: latent-space representations of points with the upper bound on the number of 3-lepton events from $\widetilde{\chi}_1^\pm \widetilde{\chi}_2^0$ production at 36~fb$^{-1}$ as a colour map.}
\end{figure}

Since the latent space groups EWMSSM models that have similar parameter values, we expect that it also groups models that feature similar production rates at the LHC. 
Searches for the electroweak sector of the MSSM regularly focus on $\widetilde{\chi}_1^\pm \widetilde{\chi}_2^0$ pair production in a scenario where $\widetilde{\chi}_1^\pm$ and $\widetilde{\chi}_2^0$ are dominantly wino and $\widetilde{\chi}_1^0$ is mostly bino, i.e.\ a scenario where $|M_1|$ < $M_2$ < $|\mu|$.
The wino assumption gives a relatively high $\widetilde{\chi}_1^\pm \widetilde{\chi}_2^0$ production cross section (compared to other production channels) and the decays to the bino $\widetilde{\chi}_1^0$ have a high likelihood of producing 3-lepton events, thereby reducing the SM-background contamination. It is therefore interesting to plot the expected number of $\widetilde{\chi}_1^\pm \widetilde{\chi}_2^0$ events as a function of the two latent-space variables (Figure~\ref{fig:latent_x1n2}).
Note that the quoted number of expected events provides a wildly optimistic view of the potential LHC reach -- no consideration is given to the potential SM background, there may be decay products with low transverse momentum that are not efficiently reconstructed, etc. Nevertheless, it is useful to determine which models are not excluded by ATLAS and CMS searches because the production rate of electroweakinos is too low. These latter models will not be good targets for further optimisation.

We may also calculate an upper bound on the number of $3$ lepton events that will be produced, by factoring in the branching ratios of $\widetilde{\chi}_1^\pm \to \widetilde{\chi}_1^0(W^{\pm(*)} \to) l^\pm \nu_l$ and $\widetilde{\chi}_2^0 \to \widetilde{\chi}_1^0(Z^{(*)} \to) ll$, where $l$ is an electron or a muon.\footnote{The cross sections used in this section are the leading-order cross sections obtained from \texttt{Prospino~2.1}~\cite{Beenakker:1996ed}, whereas the branching fractions are obtained from \texttt{SUSY-HIT~1.5}~\cite{Djouadi:2006bz}.} This allows us to explore models that conform to the assumed simplified model used to optimise the ATLAS and CMS analyses, but which may have kinematics that depart from the models that are used for search optimisation. To make sure that the leptons will have a large-enough transverse-momentum to be reconstructed in the LHC detectors, we reject models that have $m_{\widetilde{\chi}_1^\pm} - m_{\widetilde{\chi}_1^0} < 15$~GeV and $m_{\widetilde{\chi}_2^0} - m_{\widetilde{\chi}_1^0} < 15$~GeV.  The right-hand panel of Figure~\ref{fig:latent_x1n2} displays the upper bound on the number of 3-lepton events at 36~fb$^{-1}$ as a colour map in the $\vec{\theta}_{\text{2D}}$ plane. Several regions in the plane contain models with high numbers of events that are in principle ``discoverable'' in the used dataset.

\begin{figure}[t]
	\centering
	\begin{subfigure}{0.49\textwidth}
		\includegraphics[width=\textwidth]{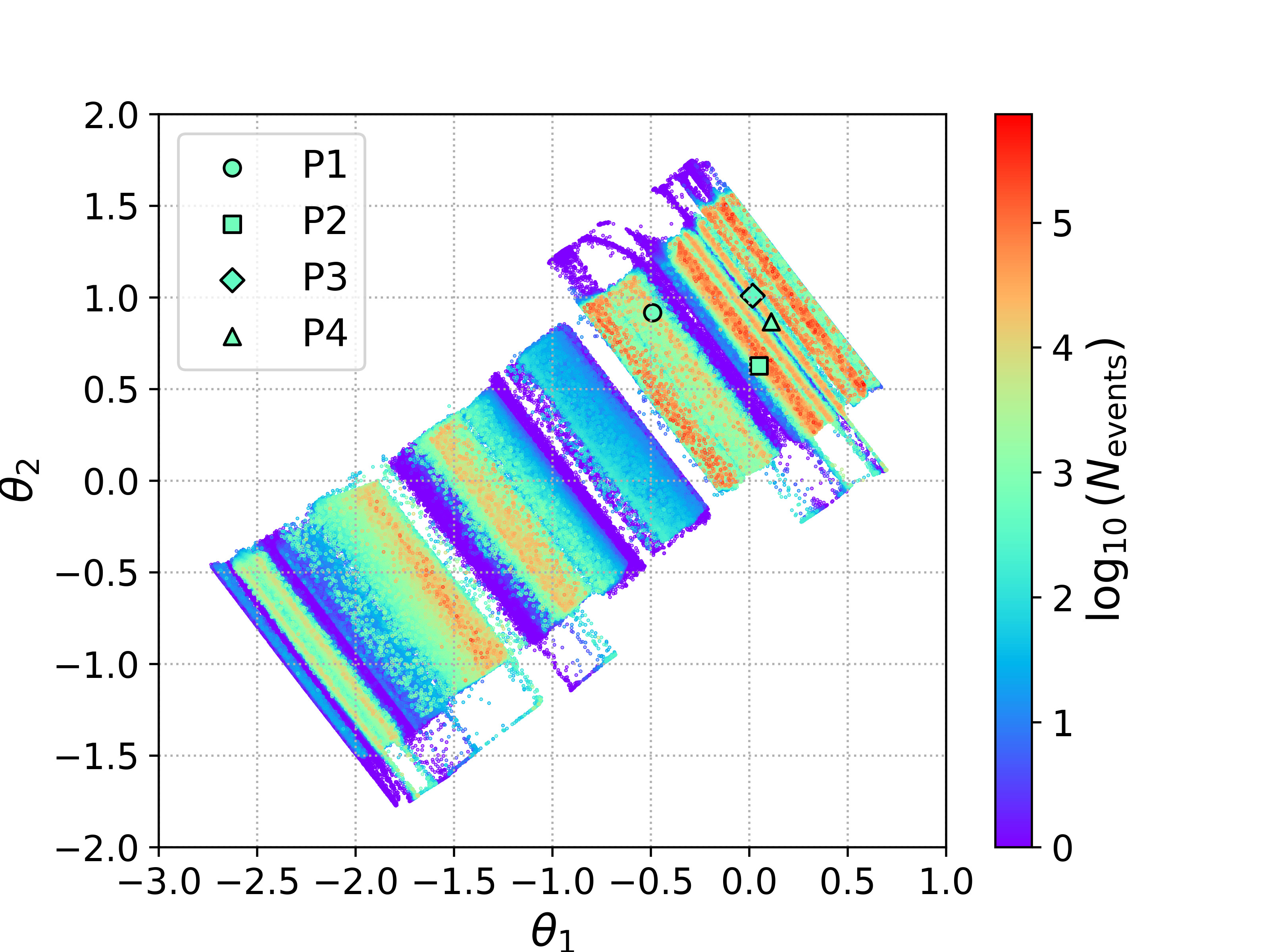}
	\end{subfigure}
	\hfill
	\begin{subfigure}{0.49\textwidth}
		\includegraphics[width=\textwidth]{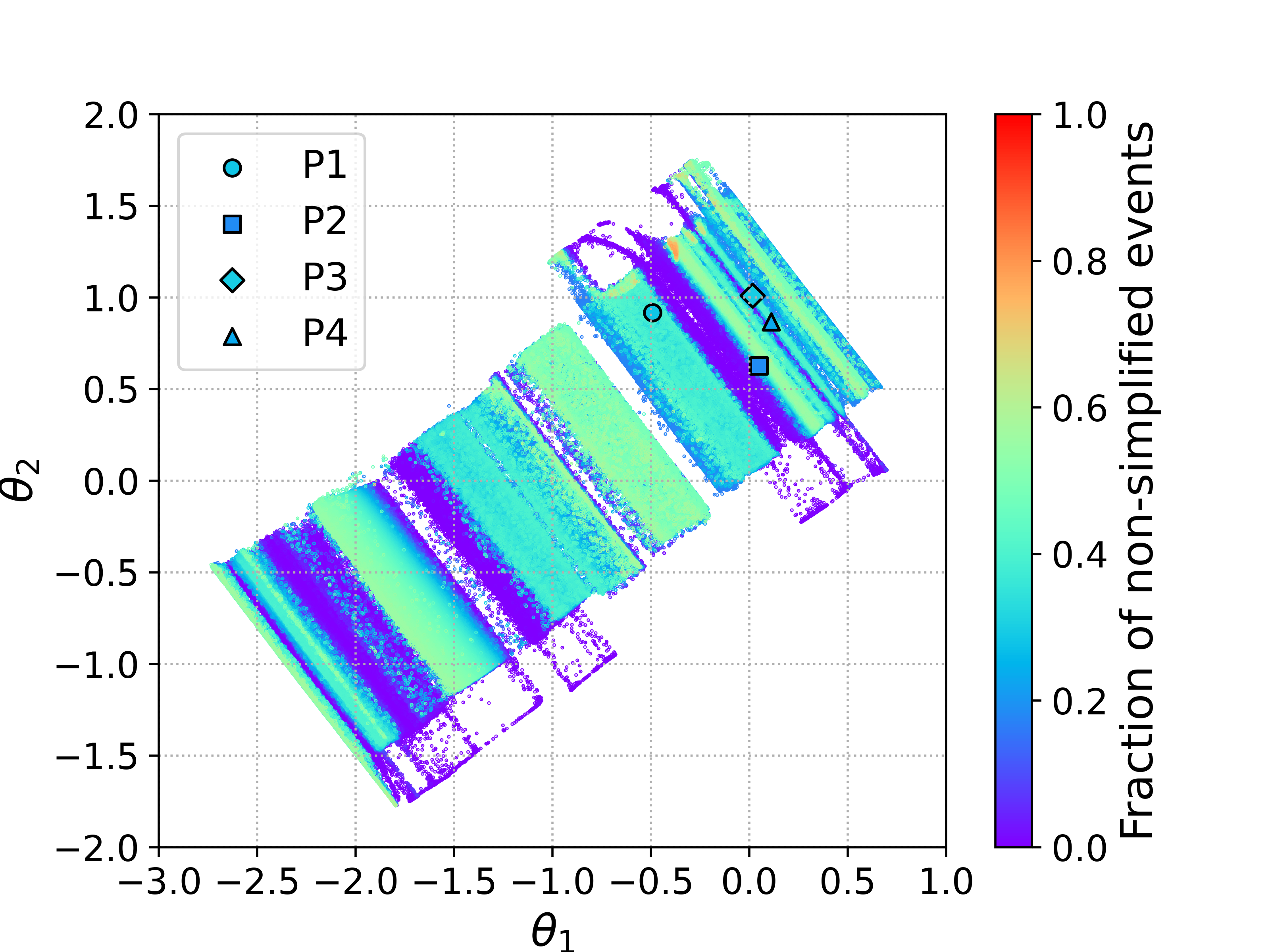}
	\end{subfigure}
	\hfill
	\caption{\label{fig:non-simplified} The number (left) and fraction (Eq.~\eqref{eq:non-simp}) (right) of non-simplified events at 36 fb$^{-1}$ in the $\vec{\theta}_{\text{2D}}$ plane. }
\end{figure}

Given that most LHC electroweakino searches are optimised on a simplified model, a particularly interesting question to ask is ``which unexcluded models exhibit behaviour that departs from the simplified model?''. There are various ways to explore this, but one simple way forward is to examine the proportion of the production cross section for each model that does not arise from the production processes typically assumed in the simplified models. To define a ``non-simplified'' process, we sum the cross sections of all production processes, but exclude the ``simplified'' and commonly used $\widetilde{\chi}_1^0$ $\widetilde{\chi}_1^\pm$, $\widetilde{\chi}^{\pm}_1$ $\widetilde{\chi}^{\mp}_1$, and $\widetilde{\chi}_2^0$ $\widetilde{\chi}_1^\pm$ pair production processes. In addition we exclude the production of $\widetilde{\chi}_1^0$ $\widetilde{\chi}_1^0$, as monojet searches are known to be weakly constraining on SUSY models. Figure~\ref{fig:non-simplified} shows the number of non-simplified events, as well as the relative fraction of non-simplified events defined as 
\begin{equation}
\label{eq:non-simp}
    \frac{\sum_i \sigma_i - (\sigma_{\widetilde{\chi}_1^0\widetilde{\chi}_1^\pm} + \sigma_{\widetilde{\chi}_1^\pm\widetilde{\chi}_1^\mp} + \sigma_{\widetilde{\chi}_2^0\widetilde{\chi}_1^\pm} + \sigma_{\widetilde{\chi}_1^0\widetilde{\chi}_1^0})}{\sum_i \sigma_i},
\end{equation}
where $i$ sums over all production cross sections. Choosing benchmark points in these regions will prioritise SUSY scenarios that are most dissimilar to those currently being used for analysis optimisation.

\section{Optimisation of analyses in the latent space plane}
\label{sec:optimisation}
In the previous section, we identified various interesting regions that should in principle be discoverable at the LHC, and which slipped through the net of the searches included in the original global fit. The invertible nature of the map from the EWMSSM parameters to the latent space plane now allows us to choose benchmark points for optimisation in the $\vec{\theta}_{\text{2D}}$ plane, and design analyses that have sensitivity to them. To demonstrate such an approach, we select four benchmark points and perform a simple search analysis showing that all of them can be excluded by a single signal region. To select useful benchmark points from the GAMBIT EWMSSM results, some care needs to be taken. Since the GAMBIT global fit represents a statistical combination of a large number of LHC searches, the surviving EWMSSM parameter points fall into two different categories: 
The first category is EWMSSM points that avoid exclusion because none of the included LHC searches are sufficiently sensitive to them. The second category is EWMSSM points that partially fit data excesses in some of the searches, thus offsetting the likelihood penalty coming from other searches that could otherwise exclude these scenarios. In the following we are only interested in the first category of EWMSSM scenarios. To identify such parameter points, we apply a number of selections to the points in the GAMBIT results:
\begin{itemize}
    \item Points $p$ must have a total log-likelihood $\ln \mathcal{L}_{\text{p}}$ satisfying $\ln \mathcal{L}_{\text{p}} > -3.09$. This is the condition that the points must be allowed at the 2$\sigma$ level in the GAMBIT analysis.\footnote{The likelihood function used in the GAMBIT study is normalised such that $\ln \mathcal{L}_{\text{p}} = 0$ corresponds to a case where the EWMMSM point gives an equally good fit to the combined set of search results as the background-only prediction does -- see Ref.\ \cite{GAMBIT:2018gjo} for further details.}
    \item Considering the log-likelihood from each individual signal region, points should not be excluded by any signal region at the $1\sigma$ level, and also should not fit any excess in the data at the same level. This ensures that the points do not belong to the second category of surviving scenarios above, where a fit to excesses in some signal regions compensate for tensions with other signal regions.
    \item We require $|\mu|<500$~GeV and $M_2<500$~GeV, which ensures that none of the winos and higgsinos are  completely decoupled. This means the EWMSSM points will have a reasonably light spectrum that a) does not resemble the ATLAS and CMS simplified models and b) gives a sizeable total production cross section.
\end{itemize}

From the set of points that survives these selections we pick four benchmark points P1--P4, covering both of the cases $|\mu|<M_2$ and $M_2<|\mu|$.
The parameters for each benchmark point are given in Table~\ref{tab:summary}, along with the electroweakino masses, the production cross sections for the most dominant channels as obtained from \texttt{MadGraph5\_aMC@NLO 2.6.3}~\cite{Alwall:2014hca} using the setup explained below, the total production cross section, and the number of events expected in 36~fb$^{-1}$ of 13 TeV collision data before any kinematic selection. 
Note that for point P2, all six of the electroweakinos are reasonably light, while for points P1, P2 and P4 there is one decoupled neutralino which is bino-dominated. For all four points, the mass difference between the
lightest and the heaviest ``active'' electroweakino (disregarding the decoupled
$\widetilde{\chi}_4^0$ for points P1, P2 and P4) is always less than 125 GeV. This configuration will result in a rich phenomenology of possible production modes and decay
chains, involving some mix of decays with on-shell $W$/$Z$ bosons and decays with off-shell $W$, $Z$ and Higgs bosons. 
As can be seen from the table, the simplified model process $\widetilde{\chi}_1^\pm\widetilde{\chi}_2^0$ often has a slightly smaller cross section than other production channels, including those with more complex decay chains (i.e.~$pp\to\widetilde{\chi}_2^\pm\widetilde{\chi}_3^0$). 

\begin{table}[t]
    \centering
	\begin{tabular}{|c|cccc|}
		\hline
	    Parameter & P1 & P2 & P3 & P4 \\
	    \hline
	    $M_1$ (GeV) & 1781 & -202 & -1336 & 1291 \\
	    $M_2$ (GeV) & 202 & 184  & 228 &  255 \\
	    $\mu$ (GeV)  & -204 & -230 & -203 & -226 \\
	    $\tan\beta$ & 36 & 46  & 17 & 27  \\
	    \hline
	    $m_{\widetilde{\chi}_1^0}$ (GeV) 
	    & 164 & 167  & 175 & 198\\
	    $m_{\widetilde{\chi}_2^0}$ (GeV)
	    & 219& 186 & 216& 241 \\
	    $m_{\widetilde{\chi}_3^0}$ (GeV)& 273 & 261 & 289 & 316\\
	    $m_{\widetilde{\chi}_4^0}$ (GeV) & 1770 & 280 & 1329 & 1285 \\ $m_{\widetilde{\chi}_1^\pm}$ (GeV)& 168& 168& 180 & 203  \\
	    $m_{\widetilde{\chi}_2^\pm}$  (GeV) & 278& 284& 292 & 319 \\
	    \hline
	    $\sigma_{\widetilde{\chi}_1^\pm\widetilde{\chi}_1^0}$ (fb)   & $3180.0$ & $4047.0$ & $2130.0$ & $1382.0$ \\
	    $\sigma_{\widetilde{\chi}_1^\pm\widetilde{\chi}_2^0}$ (fb)   & $503.1$  & $103.7$      & $582.5$  & $397.5$\\
	    $\sigma_{\widetilde{\chi}_1^\pm\widetilde{\chi}_3^0}$ (fb)   & $171.2$  & $165.8$  & $118.8$      & $81.3$\\
	    $\sigma_{\widetilde{\chi}_1^\pm\widetilde{\chi}_3^0}$ (fb)   & $<1$     & $137.4$  & $<1$      & $<1$\\
	    $\sigma_{\widetilde{\chi}_2^\pm\widetilde{\chi}_1^0}$ (fb)   & $105.1$  & $101.0$      & $79.6$  & $54.2$ \\
	    $\sigma_{\widetilde{\chi}_2^\pm\widetilde{\chi}_2^0}$ (fb)   & $184.6$  & $69.2$      & $112.0$  & $81.7$ \\
	    $\sigma_{\widetilde{\chi}_2^\pm\widetilde{\chi}_3^0}$ (fb)   & $649.8$  & $149.2$      & $625.4$  & $459.5$ \\
	    $\sigma_{\widetilde{\chi}_2^\pm\widetilde{\chi}_4^0}$ (fb)   & $<1$      & $449.8$  & $<1$      & $<1$ \\
	    $\sigma_{\widetilde{\chi}_1^0\widetilde{\chi}_2^0}$ (fb)     & $294.4$      & $59.9$      & $335.2$  & $229.7$\\
	    $\sigma_{\widetilde{\chi}_2^0\widetilde{\chi}_3^0}$ (fb)     & $91.7$      & $92.5$      & $<1$  & $39.7$\\
	    $\sigma_{\widetilde{\chi}_2^0\widetilde{\chi}_3^0}$ (fb)     & $<1$      & $36.0$      & $<1$  & $<1$\\
	    $\sigma_{\widetilde{\chi}_3^0\widetilde{\chi}_4^0}$ (fb)     & $<1$      & $76.5$      & $<1$  & $<1$\\
	    $\sigma_{\widetilde{\chi}_1^\pm\widetilde{\chi}_1^\mp}$ (fb) & $1656.0$ & $2086.0$ & $1114.0$ & $719.9$ \\
	    $\sigma_{\widetilde{\chi}_2^\pm\widetilde{\chi}_2^\mp}$ (fb) & $313.3$  & $226.3$  & $302.6$      & $220.7$\\
	    \hline
        $\sigma_{\rm tot}$ (fb) & $7149.2$  & $7800.3$ & $5400.1$  & $3666.2$ \\
        $N_{36{~\rm fb}^{-1}}$ & $2.57\cdot 10^5$ & $2.81\cdot 10^5$  & $1.95\cdot 10^5$ & $1.32\cdot 10^5$ \\
        \hline
	\end{tabular}
    \caption{Summary of the examined signal processes. We show the values of $M_1$, $M_2$, $\mu$, and $\tan\beta$, the masses of the neutralinos and charginos, the production cross sections for the most relevant channels, the SUSY production cross sections at $\sqrt{s}=13$ TeV, and the number of events expected at a luminosity of $36 \, \text{fb}^{-1}$ before any selection. Cross sections are obtained from \texttt{MadGraph5\_aMC@NLO}, where two additional jets in the hard-scattering
    matrix element are generated.}
    \label{tab:summary}
\end{table}

\subsection{Event generation}
To investigate whether these points could have been excluded with the 36~fb$^{-1}$ LHC dataset, we generate signal and background events at leading order for a $13$ TeV LHC centre of mass energy using \texttt{MadGraph5\_aMC@NLO} with the five-flavour scheme and the NNPDF PDF LO set~\cite{Ball:2014uwa}. We use variable renormalisation and factorisation scales, and all events are generated with the generator cuts set as follows:
\begin{itemize}
    \item Minimal transverse momentum of the jets $p_{T}^j > 20$~GeV, and their rapidity restricted to $|\eta^{j}| < 2.8$;
    \item Minimal transverse momentum of photons $p_{T}^\gamma > 20$~GeV and generated up to a maximum rapidity of $|\eta^{\gamma}| < 2.37$;
    \item Minimal lepton (electron $e$ and muon $\mu$) transverse momentum $p_{T}^l > 15$~GeV and with the rapidity window $|\eta^{l}| < 2.7$;
\end{itemize}
\texttt{Pythia 8.306}~\cite{Sjostrand:2014zea} is used to shower the hard process. We generate the hard process including up to two additional jets, and use MLM matching~\cite{Mangano:2006rw} to take care of the double-counting between jets produced by \texttt{MadGraph5\_aMC@NLO} and \texttt{Pythia}. The ATLAS detector is simulated using \texttt{Delphes 3.5.0}~\cite{deFavereau:2013fsa} detector simulation with a modified version of the default ATLAS detector card.\footnote{The default card was modified to ensure that low-$p_T$ leptons have a non-zero tracking efficiency.} We use the simulated LHC background processes from the dataset published and described in Ref.~\cite{Brooijmans:2020yij}, where further details on the generation of the background processes, such as the included processes and the number of generated events, can be found.

Events for each production mode of the benchmark points in Table~\ref{tab:summary} are generated in the same way as the background sample. Most importantly, since the background samples are generated with up to two additional jets in the hard-scattering
matrix element, the same is done for the signal. For each point, we drop those electroweakino production processes that would give a negligible contribution to the total number of events due to a small cross section (indicated by $<1$ in Table~\ref{tab:summary}), and generate $10^6$ MC events for each remaining process. 

\subsection{Analysis details}
In attempting to optimise prototype analyses on each of these benchmark points, one should first consider the final states that are most likely to be populated. The $\widetilde{\chi}_1^\pm\widetilde{\chi}_1^0$ and  $\widetilde{\chi}_1^+\widetilde{\chi}_1^-$ processes have the largest cross sections for each benchmark point, which naively should lead to reasonable contributions to the 0, 1 and 2 lepton final states. However, for our benchmark points, the $\widetilde{\chi}_1^\pm$--$\widetilde{\chi}_1^0$ mass splittings are below 5 GeV, and thus we instead consider more complex processes involving heavier neutralinos and charginos. The complex decay chains can more easily populate the 3 lepton final state, with extra jets plus missing energy. This final state is often used for $\widetilde{\chi}_1^\pm\widetilde{\chi}_2^0$ searches due to the small SM background. 
\begin{figure}[t]
	\centering
		\includegraphics[width=\textwidth]{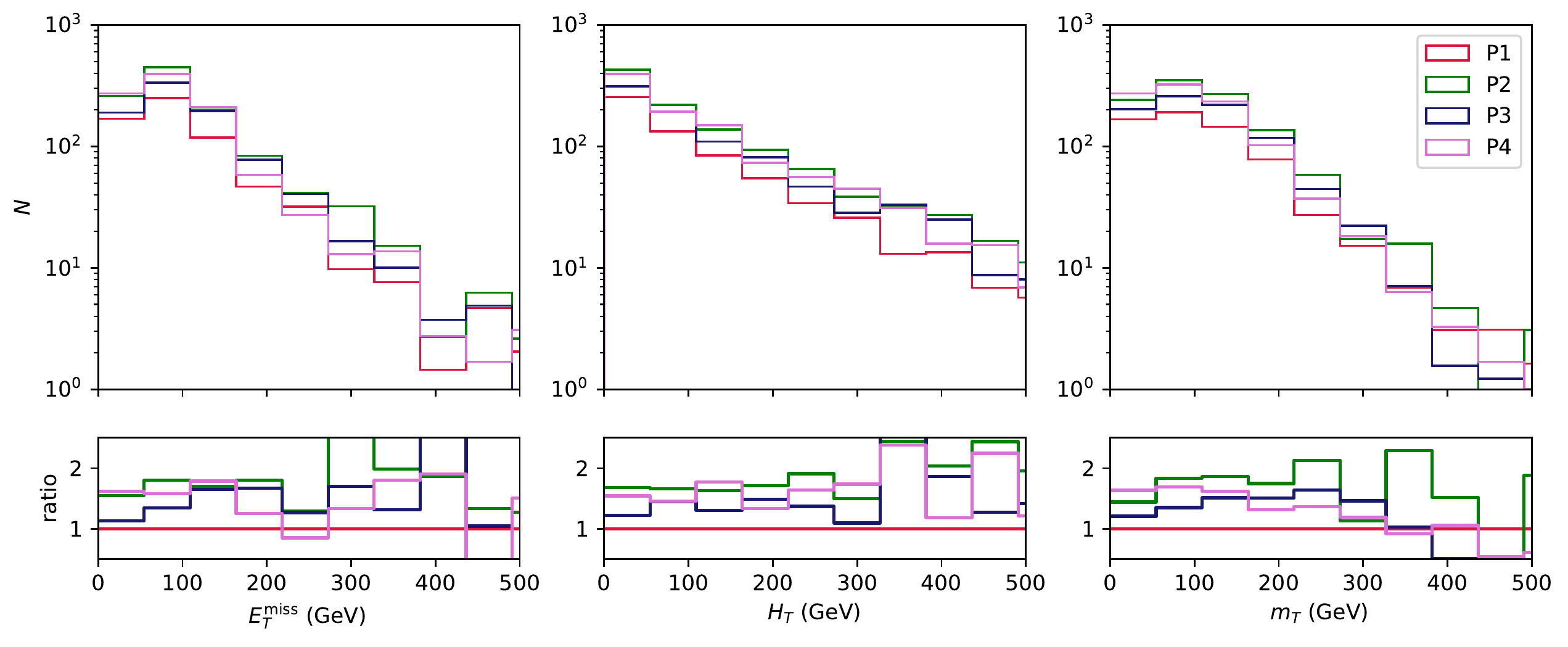}
	\caption{\label{fig:distribution_signal} Kinematic distributions ($E_T^{\rm miss}$, $H_T$ and $m_T$ defined in Eq.~\eqref{eq:mtdef})
	for the four benchmark points. The lower panel shows the ratio of the distributions to that of P1. }
	\label{fig:kindist}
\end{figure}
It turns out that all of the benchmark points can be excluded by a single signal region. This is perhaps unsurprising given that each of them involves ``non-simplified'' phenomenology with multiple options for producing gauge bosons in decay chains. We first require events to have exactly 3 leptons, and, to increase the likelihood that a pair of the leptons originates from a $Z$-boson decay, all events are required to have (at least) one same-flavour opposite-charge-sign (\textsc{sfos}) lepton pair. To ensure that the events would sit on the plateau of a suitable trigger in an LHC trigger menu, we require the transverse momentum of all of the selected leptons to exceed 20 GeV, with the leading lepton required to have a transverse momentum greater than 25 GeV. In events with 3 leptons of the same flavour, the \textsc{sfos} pair with an invariant mass closest to the $Z$-boson mass is assumed to arise from the $Z$ decay. The extra lepton is then assumed to originate from a $W$-boson decay, and we label its transverse momentum $p_T^{l_W}$. This extra lepton is used to calculate the transverse mass variable 
\begin{equation}
\label{eq:mtdef}
    m_T=\sqrt{2p_T^{l_W}E_T^{\rm miss}(1-\cos(\Delta\phi))}\,,
\end{equation}
with $E_T^{\rm miss}$ the missing transverse energy, and $\Delta\phi$ the separation in the transverse plane between the candidate lepton and $E_T^{\rm miss}$. This variable has a Jacobian peak in the $WZ$ background and drops off at $m_T \simeq m_W$ while the signal distribution is typically more flat. A $b$-jet veto is used to reduce top backgrounds. Other discriminating variables are $E_T^{\rm miss}$ itself and the scalar sum of the transverse momenta of all jets in the event ($H_T$). Our pre- and post-selection signal region selections are summarised in Table~\ref{tab:ns_analysis_cuts}. In Figure~\ref{fig:kindist} we show
the distributions for $E_T^{\rm miss}$, $H_T$ and $m_T$ for the four benchmark points after
applying the pre-selection cuts.

\begin{table}[t]
    \centering
    \begin{tabular}{c | c | c }
        & Variable & Requirement \\
        \hline
        pre-selection cuts & $n_{\text{lep}}$ & $=3$ \\
        & $n_{b-\text{jet}}$ & $=0$ \\
        & $n_{\rm sfos} $ & $=1$ \\
        & $p_{T}^{l_1}$ & $>25$ GeV \\
        & $p_{T}^{l_{2,3}}$ & $>20$ GeV \\
        \hline
        post-selection cuts &$E_T^{\rm miss}$ & $>90$  GeV\\
        &$m_{T}$ & $>20$ GeV \\
        & $H_T$ & $<125$ GeV \\
    \end{tabular}
    \caption{Summary of selection criteria for the non-simplified 3-lepton selection.}
    \label{tab:ns_analysis_cuts}
\end{table}

In Table~\ref{tab:analysis_results}, we show the number of signal and background events expected in 36 fb$^{-1}$ of data at the LHC for each of our benchmark points. We also show the binomial significance, $Z_{bi}$, calculated using the \texttt{RooStats} framework within \texttt{ROOT 6.24.02} \cite{rene_brun_2019_3895860}, with $Z_{bi}>1.64$ indicating exclusion at the 95\% confidence level. For each model, we add an assumed systematic uncertainty of 15\% in quadrature with the statistical uncertainty on the number of Monte Carlo events passing the analysis selections. Using this metric, all four signal models are able to be excluded at a 95\% confidence level.

\begin{table}[t]
	\centering
	\begin{tabular}{c | c c c | c }
		Benchmark point & $N_{sig}$ & $N_{bkg}$ & $Z_{bi}$ & Dominant production channels \\
		\hline
		\hline
		P1 & 207 & 187 & 3.8 & $\widetilde{\chi}_2^+\widetilde{\chi}_2^-$, $\widetilde{\chi}_2^{\pm}\widetilde{\chi}_3^0$, $\widetilde{\chi}_2^{\pm}\widetilde{\chi}_4^0$\\
	    P2 & 140 & 187 & 2.7 & $\widetilde{\chi}_2^+\widetilde{\chi}_2^-$,
               $\widetilde{\chi}_2^{\pm}\widetilde{\chi}_3^0$\\
	    P3 & 227 & 187 & 4.2 & $\widetilde{\chi}_2^+\widetilde{\chi}_2^-$,
               $\widetilde{\chi}_2^{\pm}\widetilde{\chi}_3^0$\\
        P4 & 218 & 188 & 4.0 & $\widetilde{\chi}_2^+\widetilde{\chi}_2^-$,
               $\widetilde{\chi}_2^{\pm}\widetilde{\chi}_3^0$\\
	\end{tabular}
	\caption{Summary of analysis results. The labels $N_{sig}$ and $N_{bkg}$ denote the number of signal and background events at 36 fb$^{-1}$ respectively. The last column shows the production 
	channels that contribute the most to the expected number of signal events. The quoted significance was calculated using the \texttt{RooStats} stats framework within \texttt{ROOT 6.24.02}. Using this metric, a $Z$ score of 1.64 corresponds to a 95\% confidence level exclusion.}
	\label{tab:analysis_results}
\end{table}

\section{Discussion of the results}
\label{sec:discussion}
We have demonstrated through a simple example how it is possible to use the latent space to identify points that were not excluded by the analyses used to obtain our list of viable models, and define a new signal region that is able to probe those points. A key element of our approach is to replace an absolute jet veto with an upper bound on the hadronic activity given by $H_T$. This passes more signal events, and allows us to apply looser selections on other kinematic variables. It is instructive to briefly compare our new signal region to the three lepton analyses included in the original GAMBIT study which showed little sensitivity to the benchmark points:\footnote{We believe that a full study of complementarity between analysis approaches is beyond the scope of this paper, since it would be more accurate to do this within the ATLAS and CMS collaborations with their dedicated software frameworks.}

\begin{itemize}
\item The ATLAS search in Ref.~\cite{ATLAS:2018ojr} featured a set of signal regions with a jet veto, and another set that allowed at least one jet. In both cases, there were tighter lower bounds on the missing transverse energy, transverse mass and lepton transverse momenta than are the case for our signal region. This made sense for the simplified models on which the ATLAS regions were optimised, but it reduces sensitivity to the models considered here. 
\item The ATLAS search in Ref.~\cite{Aaboud:2018sua} has a number of 3 lepton regions that use recursive jigsaw variables on top of a pre-selection that utlises selections on the lepton transverse momenta and the jet and $b$-jet multiplicity~\cite{Jackson:2016mfb,Jackson:2017gcy}. Three signal regions are defined with higher selections on the lepton transverse momenta than those considered here, and a requirement of low jet activity. Tighter lower bounds than those considered here are applied to transverse mass and other mass-scale variables. An extra signal region is defined specifically for the case of compressed mass spectra, in which jets are expected to arise from ISR rather than sparticle decay. The transverse mass cut is substantially higher than that considered here, and the other selections are designed to enhance sensitivity to the ISR scenario, rather than having multiple soft-ish jets produced in complex sparticle decay chains.
\item The CMS search in Ref.~\cite{CMS-PAS-SUS-16-039} is designed to target events arising from the $\widetilde{\chi}_1^\pm\widetilde{\chi}_2^0$ simplified model, in final states with 2, 3 or 4 leptons and little hadronic activity. Events are required to have either no jets or 1 jet, and the search is performed in a large number of signal region ``bins'' characterised by variables such as $m_{\text{T}}$, the transverse momentum and invariant mass of the opposite-sign same-flavour lepton pair system (where applicable), the missing transverse momentum, and different variants of the stransverse mass. To aid reinterpretations, the CMS paper also contains results for eight ``aggregate'' signal regions. These were used in the original GAMBIT study, since covariance information for the background predictions was not yet available and this made it impossible to properly utilise the finely binned signal regions. The aggregate regions have tight selections on the missing transverse momentum compared to that used in our signal region. We remain open to the possibility that our benchmark points might be excluded by a detailed treatment of the finely-binned version of the CMS analysis, though it is also possible that the requirement of low jet multiplicity removes sensitivity to them.
\end{itemize}
There are a variety of extensions via which our approach can widen the phenomenology of explored models at the LHC. For example, the invertible nature of the map from the parent model to the latent space plane means that it is possible to cover the latter with a grid of benchmark points, instead of using a simplified model plane. This has more complex behaviour than the simplified model case, since traversing the latent space plane leads to different dominant sparticle production and decay modes, which require qualitatively different analysis strategies (including changes in the targeted final state). However, our results have shown that small distances in the latent space plane cover similar models (which is expected from the ordering enforced by the VAE), and hence one could still proceed by breaking up the latent space plane into different regions, and defining signal regions to obtain maximum coverage. One can also easily visualise which regions of the plane are cross section limited, and hence undetectable in the near future. 

It would be equally straightforward to start from global fit results of other parent models and define low-dimensional latent space representations of them. In the context of SUSY searches, one could define fairly inclusive subsets of the MSSM parameters as parent models, and reduce them to 2D or 3D latent spaces which can be covered by benchmark points for optimisation. The only limitation is that of the VAE itself, which will struggle to adequately perform dimensional reduction once the disparity between the parent and latent space dimensions grows too large. Nevertheless, our results raise the prospect of carefully defining relatively low dimensional models that can be explored using global fits, before dimensionally-reducing the results for LHC optimisation. The results may not generate aesthetically-clean limits on sparticle masses given assumed branching ratios, but they are likely to cover more of the possible options for SUSY phenomenology than analyses optimised on simplified models.

Finally, it is worth noting that a more systematic exploration of \emph{simple} SUSY models, as defined above, would yield greater insights into the deficiencies of current simplified models. To give a concrete example, we have seen that a reasonable proportion of the uncovered electroweakino parameter space closely resembles the conventional simplified models for chargino and neutralino searches, but that there are also interesting regions where this is not the case. It would be possible to explore similar behaviour for other searches, e.g.~stop searches, by repeating the analysis with the addition of a minimal number of parameters to describe third generation coloured sparticles. This would allow us to find regions uncovered by the conventional stop and sbottom simplified models or, better still, to optimise searches on a new, low-dimensional latent space.

\section{Conclusions}
\label{sec:conclusion}
In this paper we have presented a variational autoencoder that is able to map viable 4-dimensional EWMSSM models from a global fit to 2-dimensional latent space representations and back again. Using this tool, we have examined the latent space and identified interesting regions which are within reach and yet were not excluded by the searches used to determine the list of viable models. The latent space plane can be used to visualise useful quantities (such as the number of events expected in a given final state), explore which regions of the viable parameter space depart from simplified model assumptions, and define benchmark models that can be used to train future analyses. As an example of the latter, we picked four benchmark points that depart significantly from the dominant simplified model used for EWMSSM searches, and showed that they can be excluded by a new set of signal region selections. 

The use of a variational autoencoder to compress model parameters to a 2D plane, combined with global fit results, allows one to optimise analyses on a broader set of phenomenological options than is covered by conventional simplified models. This raises the prospect of enhancing the discovery potential of LHC searches for supersymmetry in the near future, and the technique is easily extendable to non-SUSY scenarios. Supplementary data for this study is available via Zenodo \cite{Zenodo_dim_red_misc}.

\section*{Acknowledgements}
We thank Andy Buckley, Will Handley, Gregory Martinez, Are Raklev and Pat Scott for discussions and contributions during the early stages of this work. 
MW and AL are supported by the ARC Discovery Project DP180102209 and the ARC Centre of Excellence CE200100008. MvB is supported
by a Royal Society Research Professorship (RP$\backslash$R1$\backslash$180112),
and by the Science and Technology Facilities Council (STFC) under
grant ST/T000864/1. AK is supported by the Research Council of Norway FRIPRO grant 323985.

\bibliographystyle{JHEP}
\bibliography{bibliography}

\end{document}